# The Influence of Frequency Containment Reserve Flexibilization on the Economics of Electric Vehicle Fleet Operation

Jan Figgener[*a,b,c], Benedikt Tepe[d], Fabian Rücker[a,b,c], Ilka Schoeneberger[a,b,c], Christopher Hecht[a,b,c], Andreas Jossen[d], and Dirk Uwe Sauer[a,b,c,e]

[a] *Institute for Power Electronics and Electrical Drives (ISEA), RWTH Aachen University, Germany*

[b] *Institute for Power Generation and Storage Systems (PGS), E.ON ERC, RWTH Aachen University, Germany*

[c] *Juelich Aachen Research Alliance, JARA-Energy, Germany*

[d] *Institute for Electrical Energy Storage Technology, TU Munich, Munich*

[e] *Forschungszentrum Jülich GmbH, Institute of Energy and Climate Research Helmholtz-Institute Münster: Ionics in Energy Storage (IEK-12)), D-52425 Jülich, Germany*

[*] *Corresponding author at Institute for Power Electronics and Electrical Drives (ISEA)*

*Jaegerstrasse 17/19, D-52066 Aachen, Germany. Tel.: +49 241 80-49413*

*E-Mail address: jan.figgener@isea.rwth-aachen.de*

***Abstract* – In recent years, the market for frequency containment reserve (FCR) has become a relevant source of revenue for stationary battery storage systems in Germany. During this period, prices for FCR have decreased, while the market has become increasingly flexible with shorter service periods and lower minimum power requirements. This flexibility makes the market attractive for pools of electric vehicles (EVs). Their idle times are now often longer than FCR service periods, providing the opportunity to earn additional revenue. In this paper, multi-year measurement data from 22 commercial EVs are used to develop a simulation model for FCR commercialization. In addition, the driving logbooks of more than 460 vehicles from different commercial fleets are analyzed. Based on our simulations, the impact of FCR flexibilization on the economics of an EV pool is analyzed using the German FCR market design from 2011 to 2020. It is shown that depending on the fleet, especially the recent change in service periods from one week to four hours generates the largest increase in available pool power. Further reductions in FCR service periods will like produce minor benefits, as idle times are often longer than service periods. Overall, the increase in flexibility greatly offsets the decreasing FCR prices and leads to higher revenues for most fleets analyzed. According to our model, revenues of about 250 €/a to 400 €/a could have been achieved per EV in the German FCR market in 2020.**

## I. Introduction

This section presents the thematic overview, a summary of existing literature on vehicle-to-grid (V2G) concepts with focus on the provision of frequency containment reserve (FCR) through electric vehicles (EVs) and highlights the scientific contribution of this paper. Figure 1 provides a graphical overview of the paper. First, two databases (EV measurements and driving data) and EV master data are used for the development of a simulation model. The results of the simulation model are power capability profiles, which are the bidirectional power potential of different EV pools. The profiles are used together with historical FCR price data within a calculator to estimate FCR revenues for different market designs. We show that the increasing flexibility of FCR service periods overcompensates for falling FCR prices and that revenues increase, albeit at a low level.

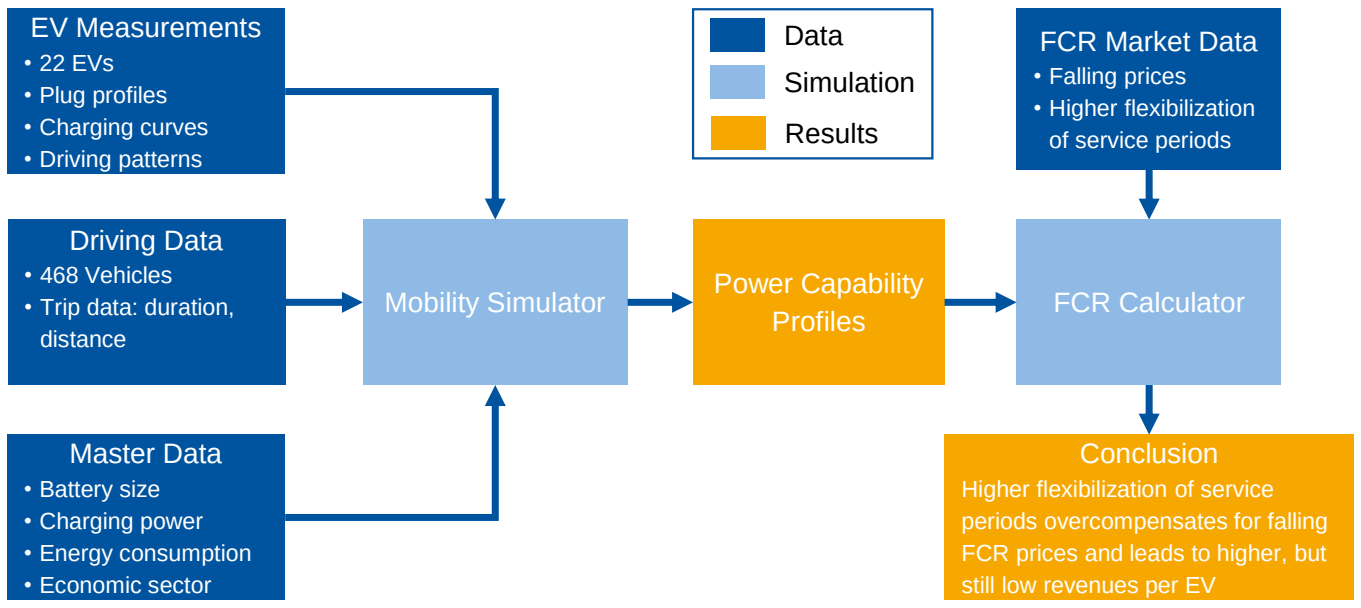

*Figure 1: Graphical abstract.*

### I.A. Motivation and contribution

The increasing integration of volatile renewable energy generation and flexible energy storage has led to an increasing flexibilization of spot and grid service markets in recent years. The term flexibilization refers on the one hand to the reduction of service periods for the respective market and on the other hand to the introduction of the required minimum power bids to participate in the market. Battery storage systems (BSSs), in particular stationary large-scale BSSs, already have a significant share of the global market for FCR, the fastest type of frequency control power in Germany. From this market, they continue to displace conventional market participants such as large power plants. In Germany, the share of stationary large-scale BSSs in the FCR market was about two thirds in 2019 [1,2]. However, these BSS have the problem that they are refinanced exclusively from FCR revenue. The ensuing price competition in order to win a contract in the bidding process has led to a 50% decrease in FCR prices from 2015 to 2020 and made the market increasingly unattractive as the main application for these large-scale BSSs [1]. For this reason, so-called multi-use concepts for BSSs are a promising way of generating revenue from various applications [3]. Decentralized BSSs can participate in virtual power plants at times when they do not fulfill their primary use. Such a pool can consist of BSSs of all types and includes stationary as well as mobile BSSs in the form of EVs. EVs, in particular, have a great potential of free battery capacities that are not used for mobility due to idle times of more than 95% [4,5]. Especially commercially operated EVs often have short and regular (and thus plannable) distances and driving patterns that allow the provision of grid services [6]. Therefore, the increasing flexibility of the FCR market (decrease of minimum power bid from 5 MW to 1 MW and shortening of service periods from one month to 4 hours in recent years) makes this market attractive for the fluctuating power of EVs in multi-use concepts. This paper analyses the influence of FCR flexibilization on the profitability of commercially operated EVs in the use case of the German market design, which should be representative for a large region of Central Europe, since the FCR market of Germany, Belgium, the Netherlands, France, Switzerland, and Austria is coupled and the price varies only slightly between the countries [7]. Even though there is a variety of scientific publications on FCR and EVs, none of the literature focuses on the recent FCR flexibilization and its impact on the economics of EV fleets potential in this market. While other publications focus mainly on optimized operation strategies and multi-use in a fixed market design, this paper analyzes the influence that the market changes themselves have on the economics of an EV fleet. This analysis has so far not been examined (see section I.B). We contribute with our paper to fill this gap. The key research questions answered are:

1. How much FCR power can commercial EV fleets offer over different time periods? (section III.A)
2. How much money can EVs expect to earn through FCR? (section III.B)
3. What had greater impact on the economics of providing FCR through EVs over the last years: the increasing market flexibilization or the decreasing FCR prices? (section III.B)

### I.B. Literature review and differentiation

There are many publications on the provision of frequency regulation by stationary BSSs or EVs, each with a different focus and data. We divide our literature research into the areas (1) provision of FCR using BSSs and combination of applications (see Table 1), (2) simulation and optimization of frequency regulation using EVs (see Table 2) and (3) demonstrations, experiments and field tests of frequency regulation using EVs (see both Appendix, Table 9 and Table 10). We classify the sources according to Table 1 into respective focal points, which we discuss individually in the following.

(1) The provision of FCR with large-scale BSSs has been investigated and shown to be possibly profitable, depending on the energy-to-power ratio (EPR) and specific market conditions [8–10]. Especially the so-called “30-minutes” criterion when providing FCR with batteries in the German regulatory market had been determined as a crucial burden for providers [9,10]. This criterion required that the maximum offered bidirectional FCR power must be able to be provided for at least 30 minutes at any time within the respective service period. In 2019, the German federal network agency (FNA) obliged the transmission system operators (TSOs) to apply the 15-minute criterion instead of the 30-minutes criterion for BSSs making the market more attractive to batteries due to smaller EPRs (see section II.F) [11]. Multi-use, the combination of different storage applications, has been studied for years [3,12–15]. The main results of these studies are that the combination of different storage products increases the economic attractiveness, but that there are regulatory hurdles to overcome [12–14]. In this context, Englberger et al. published an open-source tool simulating multi-use including an economical and a technical analysis [15]. However, all mentioned publications focus on stationary applications and do not evaluate the flexibilization of the FCR market, on which this paper focuses.

(2) The concept of Vehicle-to-Grid (V2G) was first introduced by Kempton et al. [16,17]. Since then, the concept has been studied extensively. Table 2 provides publications, in which the provision of FCR with EVs has been simulated and optimized. The provision of different frequency regulation products by EVs offers economic potential for the owner and the aggregator [18–21]. Furthermore, the grid can benefit from it [19]. Moreover, control algorithms for aggregators have been developed and bidding strategies optimized [22,23]. V2G can be carried out using unidirectional and bidirectional chargers. Despite their momentarily higher purchase costs, bidirectional chargers offer higher economic potential [24]. In addition, simulations show that the energy throughput is increased by the provision of frequency regulation, which might lead to increased battery degradation [20,21]. However, none of the mentioned publications evaluate the impact of recent FCR flexibilization.

**Second-life use vs. dual use**

The concept of **second-life use** means that after years of use for mobility, vehicle batteries are removed from EVs and integrated into other applications like stationary BSSs in order to provide grid services or trade energy [25–27]. However, batteries contained in EVs can also be used to provide these V2G services within their first life, if they are temporarily not required for their primary use, namely mobility. We therefore call this concept **"dual use".**

(3) Several demonstration projects have been and are being carried out to investigate the provision of frequency regulation with EVs. Appendix, Table 9, gives a selection of such projects, while Appendix, Table 10, lists scientific publications done within these projects. In 2002, the first project on frequency regulation supply with EVs was implemented in California [28,29]. It showed that EVs are capable of providing frequency regulation to the grid. The German INEES project ran from 2012 to 2015 and analyzed the provision of secondary control reserve with an EV fleet of 20 V2G-compatible vehicles [30,31]. The provision proved to be technically possible, but not profitable under the conditions prevailing at the time [30]. In addition, the impact on the distribution grid in terms of power quality and manageability was not assessed as negative [31]. Another demonstration project ran from 2013 to 2018 in California, within which 29 bidirectional EVs provided frequency regulation [32,33]. Among other things, optimization models were used to minimize operating costs and maximize revenue from ancillary services [33]. The so-called “Parker” project, within which many scientific findings were published, ran from 2016 to 2019 in Denmark [34–40]. The scientists showed that FCR supply is possible with unidirectional charging stations, but is economically much more attractive with bidirectional charging stations [35,36,39]. Furthermore, it was found that the response times and accuracies of the charge controllers are sufficient to provide control power [36,37]. Nevertheless, communication delays and measurement errors turned out to be practical obstacles [38]. Factors influencing the economic benefit of providing control power are the availability of vehicles, the charging efficiency and the operation strategy used [38–40]. In addition, an industrial project ran between 2018 and 2019 in which the provision of frequency regulation with a prequalified EV was successfully tested in Germany [41]. Another project, monitored by the German research institution FfE (Forschungsstelle für Energiewirtschaft e.V.), started in 2019 and analyses use-cases of EVs in different electricity markets [42]. To analyze the interaction between EVs, charging infrastructure and the grid, 50 EVs will be tested in the field [42,43]. The interconnection of EVs to a virtual power plant providing frequency regulation will be investigated in another industrial project until 2021 [44,45]. However, these projects focused on the demonstrations of V2G applications and did not focus on the market side as this paper does.

*Table 1: Summary of selected literature of the provision of FCR using BSSs and multi-use.*

| | Source | Date | Focus | Results |
|---|---|---|---|---|
| Providing FCR with battery storage systems | Fleer et al. [8] | 2016 | Economics of the provision of FCR using BSS based on two case studies considering FCR prices and battery aging | - BSS only profitable with a power-to-energy ratio of 1:1, not with 1:2<br>- Decreasing battery prices will increase possible profit but could lead to lower achievable revenues due to market saturation |
| | Zeh et al. [9] | 2016 | An optimal control algorithm for the operation of FCR with BSS is developed for the market conditions of 2015 | - FCR Market conditions for BSS of August 2015 (30-minute-criterion) lead to unprofitable operation |
| | Thien et al. [10] | 2017 | Operation strategy for an installed 5-MW-BSS providing FCR is developed and influencing parameters are analyzed | - Benefits could be enhanced having better market conditions such as 15-minute instead of 30-minute-criterion |
| Multi-Use | Fitzgerald et al. [12] | 2015 | Evaluation of different services BSS can provide in the US market including a meta-study and analysis of barriers | - Combination of applications increase the economic value<br>- Despite technical readiness, regulatory hurdles exist that prevent an economically profitable use of BESS |
| | Stephan et al. [13] | 2016 | Analysis of investment attractiveness of different single applications of BSS and their combination by developing a techno-economic model | - Combination of applications improves the investment attractiveness<br>- Market barriers often prevent the combination of applications |
| | Braeuer et al. [14] | 2019 | Evaluation of economics of BSS installed in German small and medium sized enterprises when combining applications of peak-shaving, FCR and arbitrage | - Individual applications are not profitable, but combination of applications are<br>- Influence of arbitrage application is small |
| | Englberger et al. [15] | 2020 | Simulation of energy storage systems serving multiple applications including an analysis of technical and economical parameters | - Application stacking more economical than single use<br>- Publication of open-source tool which combines BSS applications |

*Table 2: Summary of literature about simulation and optimization of the provision of frequency regulation using EV fleets.*

| Source | Date | Focus | Results |
|---|---|---|---|
| Tomić et al. [18] | 2007 | Analysis of two fleets of utility EV providing power for regulation services in the US. | - V2G enables potential revenue streams for EV owner in most ancillary service markets |
| Han et al. [22] | 2010 | Proposition of an aggregator pooling EV to provide frequency regulation | - Development of an optimal control strategy for EV fleet considering battery energy capacity and desired final SOC for driving purpose |
| Sortomme et al. [19] | 2012 | Development of a V2G algorithm for the scheduling of the provision of the ancillary services load regulation and spinning reserves in US markets | - Algorithm combines several ancillary services<br>- Simulations show that even though there are challenges, providing ancillary services can provide benefit for the owner, the aggregator and the grid. |
| Bessa et al. [23] | 2012 | Optimized bidding of EV fleet in day-ahead and secondary reserve Iberian market. | - Using an aggregator to optimize the bidding of energy decreases charging costs<br>- Variables like the electricity price and the maximum available power in each time interval need to be forecasted |
| Codani [24] | 2015 | Simulation of participation of 200,000 EV in FCR market in France | - Potential revenue higher with bidirectional charging compared to unidirectional charging<br>- A high number of EV might saturate FCR market |
| Hoogvliet et al. [20] | 2017 | Economic analysis of the potential revenue EV owners can raise when providing regulating power in the Netherlands | - Depending on EV type and driving pattern an EV owner can raise 120 € to 750 € per year when providing regulating and reserve power<br>- Provision will lead to higher battery energy throughputs (11% to 55%) |
| David et al. [21] | 2017 | Economic analysis of the provision of frequency regulation using EV considering battery degradation and driving requirements | - EV with highest battery capacity lead to the greatest economic benefit, as the cyclic degradation is lowest<br>- Major constraint is the power capability of the EV and the chargers<br>- Incentives should be developed to convince EV owners to provide frequency regulation |

The most important project (without focus on FCR) for this paper, “GO-ELK”, was conducted by the Institute for Power Generation and Storage Systems at RWTH Aachen University [46]. Within this project, 22 commercially operated EVs were equipped with data loggers to measure quantities such as battery voltage and battery currents during charge and trips. The logged data build the basis of many of the results shown in this paper.

## II. Methodology

This section describes the paper`s methodology. It presents the used data, the developed driving profile generator, the modelling approach, and the market for FCR in Germany.

### *II.A. Data collection*

This paper uses two databases containing the use of commercial vehicles. Table 3 compares the most important data of both databases and gives further information on calculations and assumptions.

The first database, “Measurements”, was created by the Institute of Power Generation and Storage Systems (PGS) at RWTH University. The high-resolution data (T = 1s) of commercially operated electric vehicles were measured between 2013 and 2016 within the project "Commercially operated electric vehicle fleets (GO-ELK)" [46]. In the project, four fleets of EVs were deployed in different sectors over a period of 30 months [46]. During their use, vehicle data (driving, charging and battery data) of the total of 22 EVs were recorded by data loggers in the vehicles and the charging stations. For a detailed description of data collection and adjustment, please refer to [46–49].

The second database, “Logbook”, was included within the project REM 2030 (regional eco mobility). The project was supervised among others by the Fraunhofer Institute for Systems and Innovation Research ISI and various institutes of the Karlsruhe Institute of Technology (KIT) and covered different topics of future urban mobility [50]. The goals were new innovative traffic concepts of individual mobility in order to avoid local emissions [50]. Within this project, Fraunhofer ISI collected travel data of commercial vehicles, which can be used free of charge for non-commercial purposes. The vehicles are not specifically electric vehicles. The relevant parameters of electric vehicles such as energy content of the battery and consumption are supplemented in this work (see Appendix, Table 8). The database contains over 91,000 journeys of 630 vehicles of different trades. We filtered the database so that only vehicles with a minimum number of one trip and a minimum logging duration of one week were considered. The vehicles are classified according to their economic sector (NACE criteria) [51]. The KIT classified the 630 vehicles used from database „Logbook“ into 15 economic sectors. Within the economic sector, there are further definitions of vehicle classes. The vehicles measured by us from our database “Measurements” are also divided into the respective economic sectors and supplement the large KIT data set. With 94 vehicles, the manufacturing sector has the largest number of vehicles, especially of the vehicle class "medium" (see Appendix, Figure 21 (left)). This is followed by public administration (71 vehicles) and human health service (58 vehicles), in which mainly small vehicles are used. Some clusters follow with 30-40 vehicles and there are also four clusters with less than 10 vehicles each. This should be considered when looking at the results, as the power capability profiles in these cases are only based on small sample sizes. The average duration of trip recording for most clusters is about 20 days (see Appendix, Figure 21 (right)). The analyses of our database “Measurements” (data recording periods of more than one year) show that these relatively short durations are already sufficient to reliably map daily operations in the examined sectors.

*Table 3: Available data in the two databases used in this paper.*

| | Category | Database “Measurements” (22 cars) | Database “Logbook” (468 of 630 cars) |
|---|---|---|---|
| Vehicle Data | Product | see Appendix, Table 7 (e.g. Smart ED, Renault Zoe) | see Appendix, Table 8 |
| | Vehicle class | small to medium | see Appendix, Table 8 |
| | Type of vehicle drive | electrical drive | combustion drive |
| | Industry / trade | see Appendix, Table 7 (e.g. healthcare and energy provider) | see Appendix, Table 8 |
| | Environment | urban and rural | urban and rural |
| Trip Data | Start of trip | timestamp | timestamp |
| | End of trip | timestamp | timestamp |
| | Trip consumption in kWh | ~15-25 kWh/100 km | 18.9 kWh/100 km to 27 kWh/100 km (see Table 8) |
| | Trip distance | measured in km | measured in km |
| | Distance to company at trip end | taken from charge events | measured with GPS |
| Battery Data | Max. charge power in kW | 3.7 kW to 22 kW (AC) | according to “Measurements” |
| | Battery energy in kWh | 16 kWh to 24 kWh | 19.1 kWh to 80.7 kWh (see Table 8) |
| | Charging curve | measured in laboratory | according to “Measurements” |
| | Start of charging | measured timestamp | according to “Measurements” |
| | End of charging | measured timestamp | according to “Measurements” |
| Charging Station Data | Charge power | 1.8 kW (single phase AC) to 11 kW (three phase AC) | according to “Measurements” |

*II.B. Statistical analysis*

In the following, we present a short statistical evaluation of the data used. For better comprehensibility, we present the human health service as an example from database “Measurements”. This fleet shows the same driving pattern for the whole week, which makes the daily analysis done very representative. Further analyses on database “Logbook” indicate in general similar results as the evaluation done for the “Measurement” database. All vehicles of the two databases were evaluated analogously to the evaluations presented in this section. However, as this paper does not represent a mobility study, no further evaluations are made at this point and reference is made to mobility studies such as [52–55].

Figure 2 shows the three events "plug & charge", "end of charge", and "unplug" of the Smart ED of a healthcare service. The health-care service runs in two shifts a day: 50% of all unplug events are at around 7 a.m., which is the start of the first shift. It ends at around noon and the cars are plugged and charged. The second shift runs from around 2 p.m. to 9 p.m., which is again indicated by the “plug and charge events”.

The average distance (see Figure 3) between two charging operations is about 40 km, although the maximum possible distance range of the vehicle is around 100 km. Further, the average duration between unplug und plug is about 8 h (see Figure 4). All travel durations above about 8 h can be assigned to shifts after which no charging process is initiated. This is the case for about 30% of all trips. The regular charging with 11 kW charging stations at the end of the shifts leads usually to a fully charged vehicle a short time after returning. While the average SOC after return is around 60%, this value is often 100% or near to 100% when the vehicle starts a trip (see Figure 5). The unplug events where the EV is not fully charged in the morning are mainly due to software issues with the internal SOC estimation showing values slightly below 100% although the EV does not charge any more. The small amounts of needed energy underline the potential of free capacity ranges shown in recent literature (see section I.B). Due to the relatively short distances, the vehicle consumes on average only 40% to 50% of the battery energy capacity (see Figure 6). The measurements also contain different consumptions as a function of the temperature as the EVs were measured over two years and thus within different seasons (a low temperature leads to high consumptions and vice versa).

Database “Logbook” contains only vehicles with internal combustion engines. In order to model these as fully electric vehicles, the unrecorded consumption must be estimated. The values shown in Appendix, Table 8, are used to estimate the consumption. These result from various studies, test results and manufacturer data sheets of EVs that are in the same vehicle class as the internal combustion engine vehicles. The average consumptions of the listed EVs range from 18 kWh/100 km and 25 kWh/100 km depending on the vehicle class [56]. To determine the consumption in kWh per 100 km of a trip, the consumption is distributed normally around the average consumption of the vehicle class (expected value: average consumption of all EVs in Appendix, Table 8, variance: 1 kWh / 100 km). This way, the effect of the temperature on the consumption is also statistically taken into account.

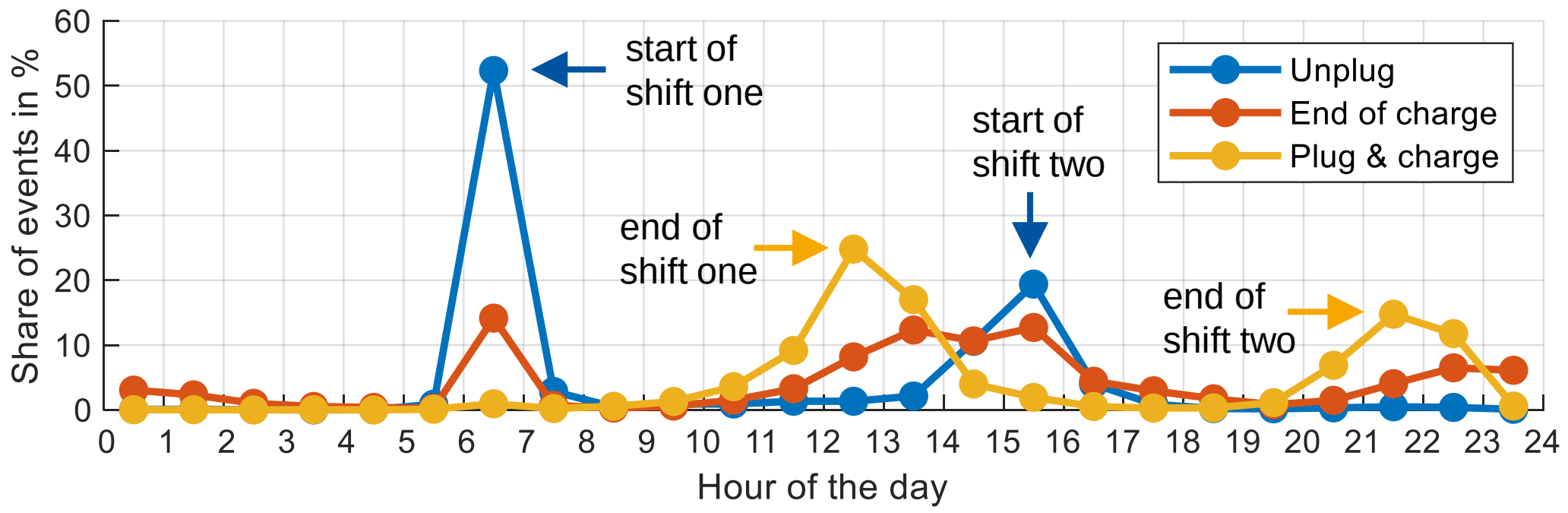

Figure 2: Trip events of EVs in human health (908 measured trips from database „Measurements“).

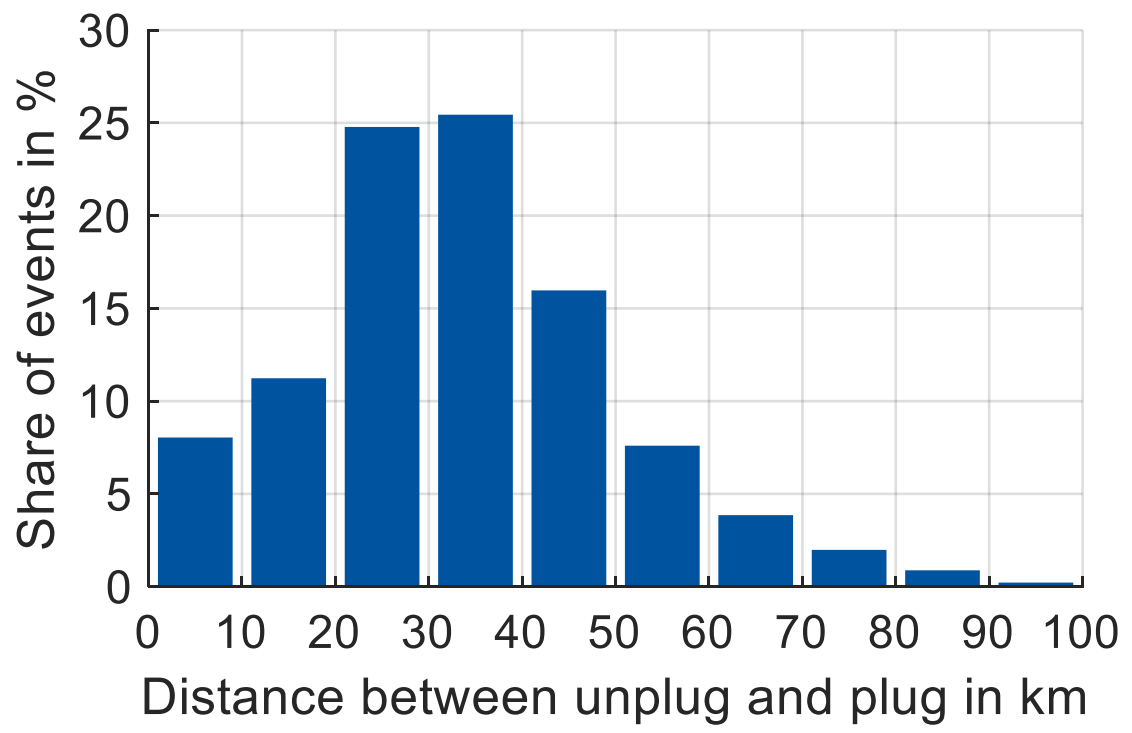

*Figure 3: Distance between unplug and plug (908 measured trips of EVs in human health from database „Measurements“).*

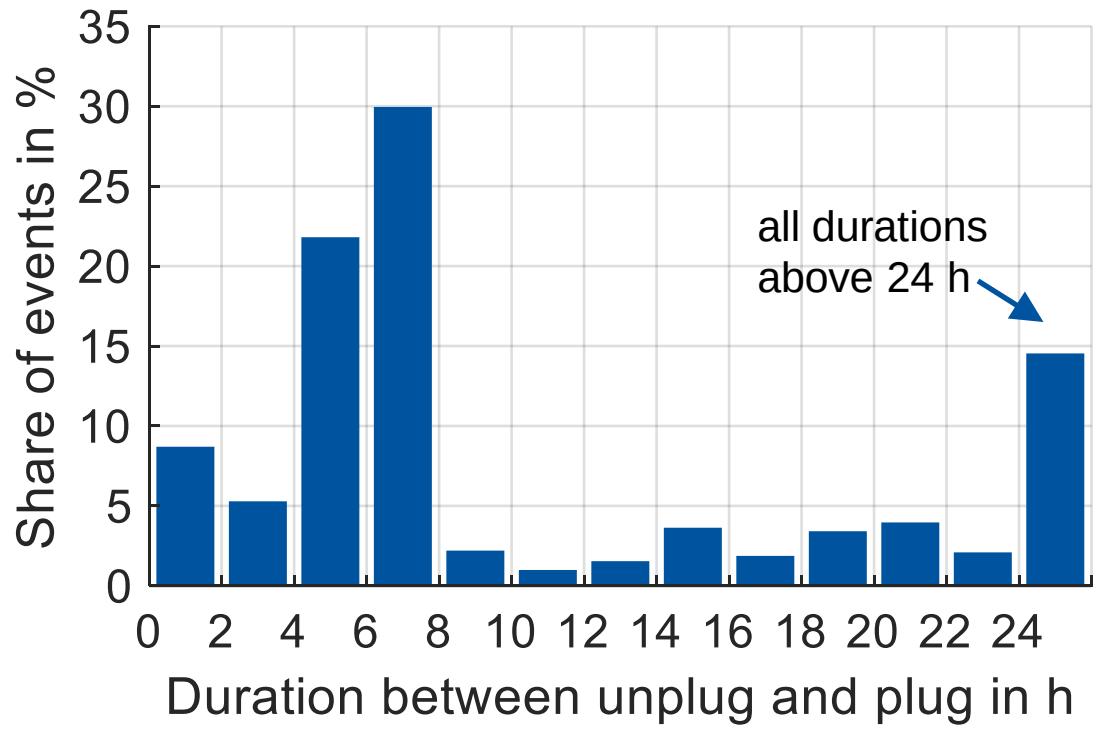

*Figure 4: Duration between unplug and plug (908 measured trips of EVs in human health from database „Measurements“).*

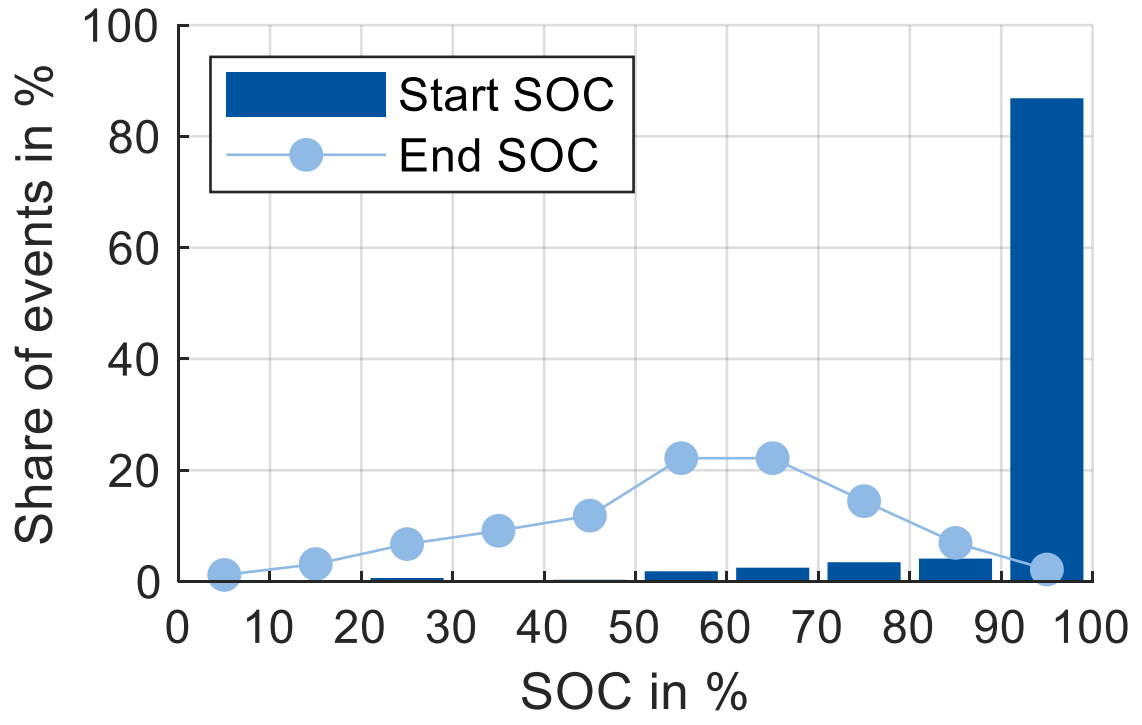

*Figure 5: SOC at unplug and plug (908 measured trips of EVs in human health from database „Measurements“).*

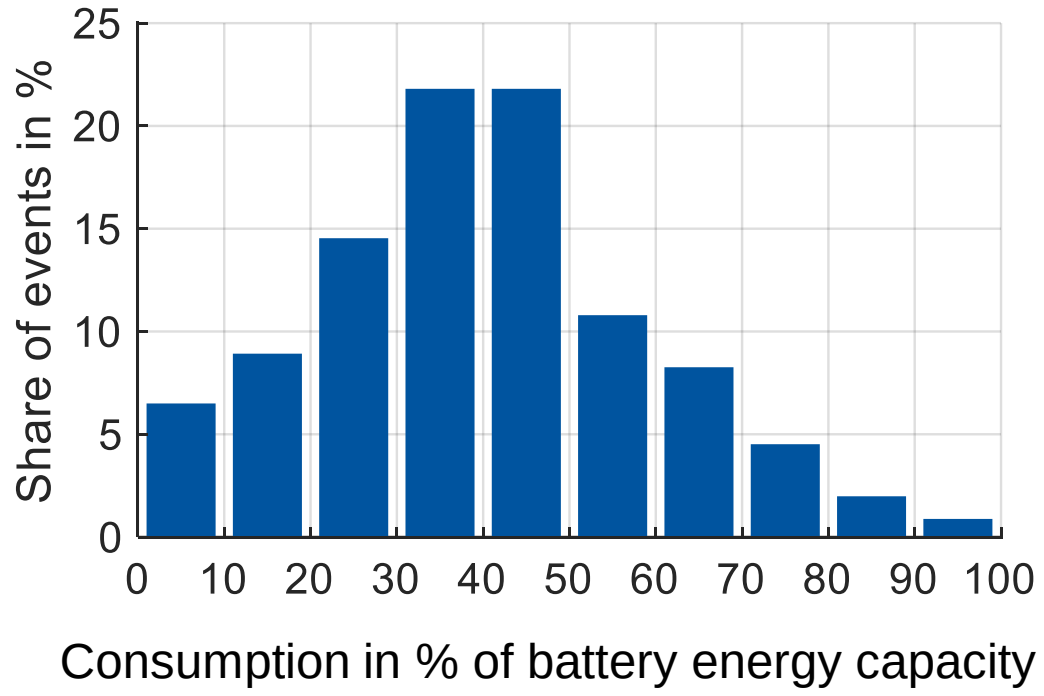

*Figure 6: Consumption between unplug and plug (908 measured trips of EVs in human health from database „Measurements“).*

*II.C. Driving profiles*

The statistical analysis is now used to generate driving profiles such as the probability of a trip start or day and time dependent distributions of distances, durations, and energy consumptions. The data is presented for a Smart ED that was part of the human health fleet in the statistical analyses shown in section II.B.

The trip-start probability $w_{start}$ of a plugged vehicle is calculated as a function of day and time. To get the trip-start probability, the number of unplug events are divided by the total number of days, when the car was connected to the charging station (see equation (1) and nomenclature in Appendix, Table 6).

$$w_{start}(t) = \frac{\sum_{n=1}^{N} trip_{\ start,n}(t)}{N} \qquad (1)$$

$$with\ trip_{start,n}(t) = \begin{cases} 1, & if\ trip\ started\ on\ day\ n\ at\ time\ t \\ 0, & otherwise \end{cases}$$

$$with\ N = number\ of\ days, where\ EV\ was\ plugged$$

The trip-start probabilities are exemplarily shown in Figure 7. The probability for starting a trip is over 60% for the first shift and is most likely between 6 a.m. and 7 a.m. During the second shift, with cumulated probability values of approx. 40%, the number of unplugs is slightly lower. Outside the shifts, the probability of plugging out from the charging station with values below 10% is relatively low. Further, the trip distances (Figure 8), durations (Figure 9), and normalized consumptions are sorted by day and time to ensure a realistic driving behavior of the analyzed vehicles. The figures show the probabilities and distribution clustered for one hour for clearer presentation. The resolution used for simulations is 15 min and thus higher (see section II).

The value distributions presented, such as distance, duration, and consumption, are also evaluated as a function of time, resulting in separate distributions for each point in time. The distances driven and the durations between plug and unplug are subject to temporal fluctuations. Both, the longest distances and durations are the start of each trip of the two shifts in the healthcare service. Especially if the EV is not plugged after the end of the first shift, the distances and durations get longer as their values also cover the second shift and vice versa.

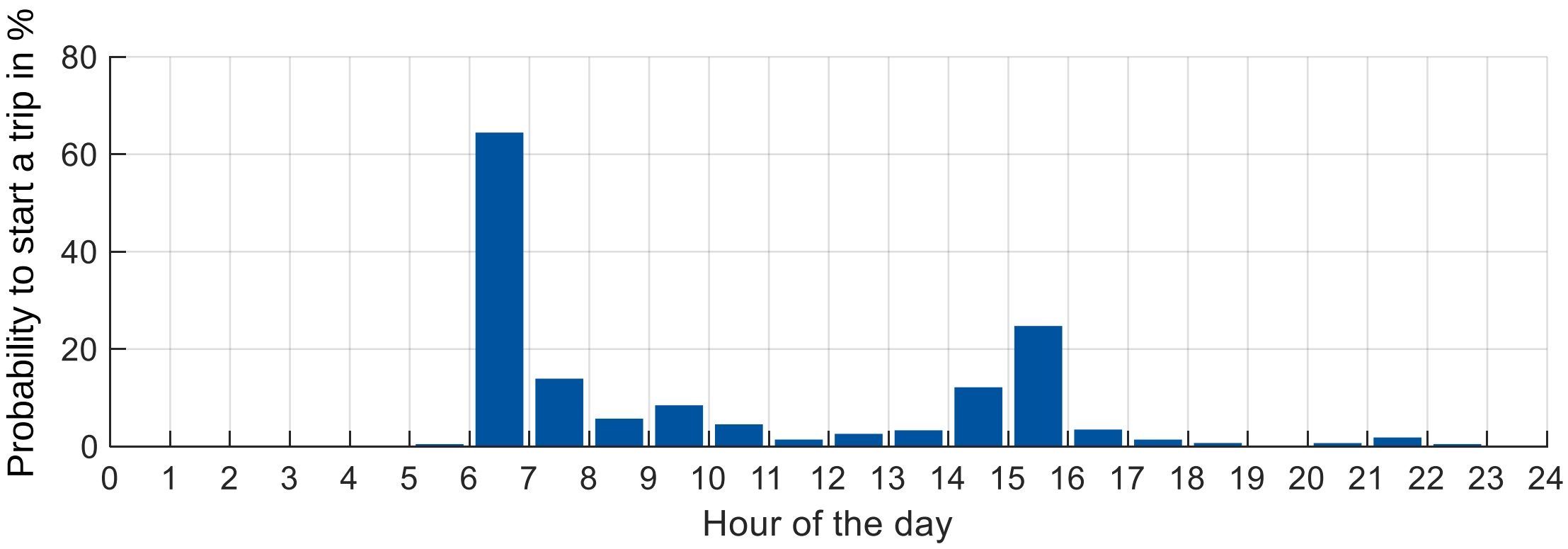

*Figure 7: Trip-start probability (Smart ED, human health).*

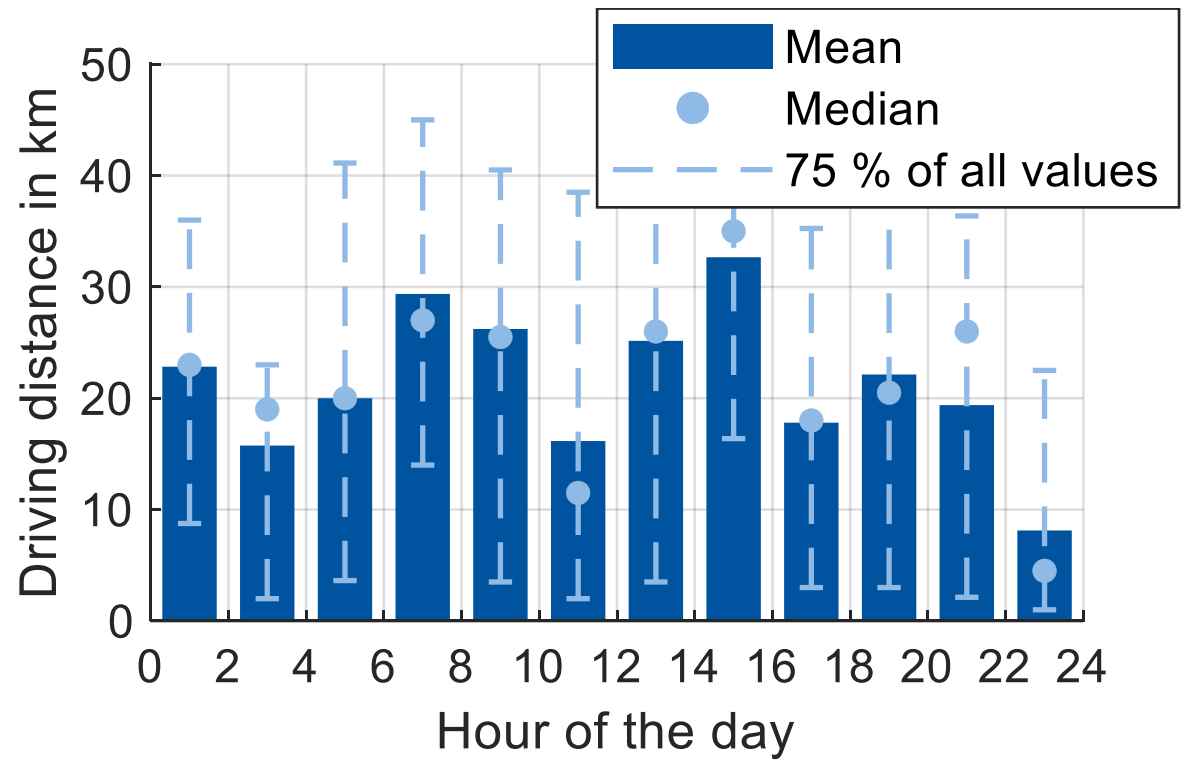

*Figure 8: Distance distributions as a function of the departure time (Smart ED, human health).*

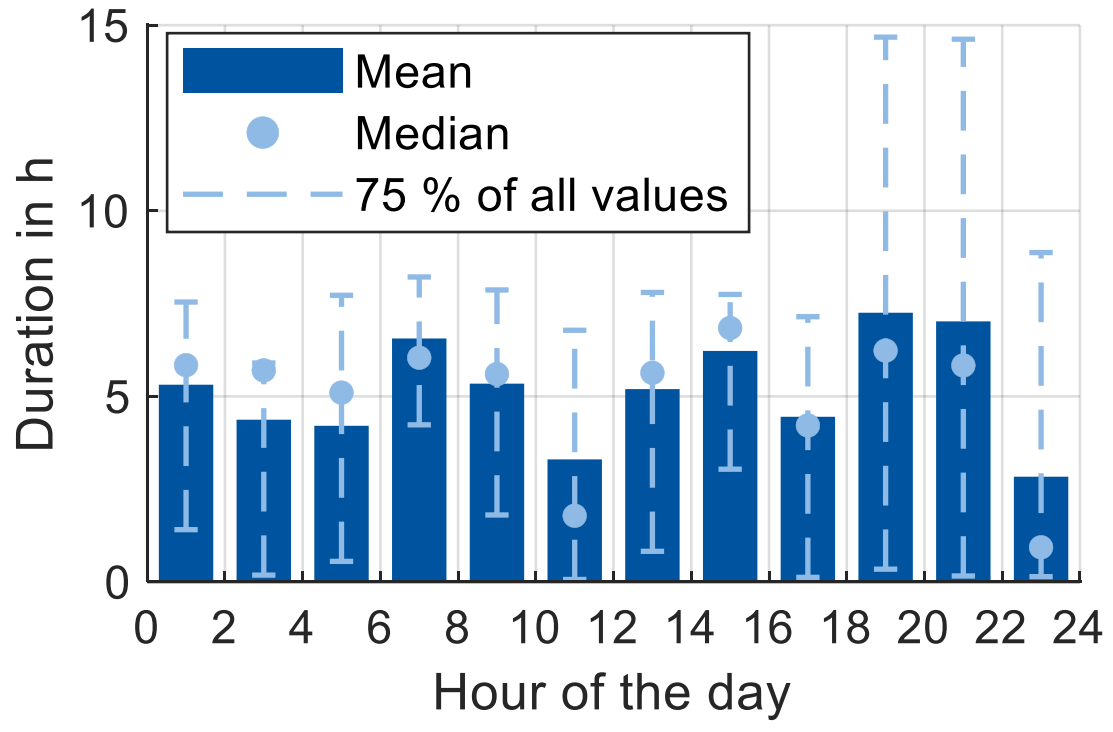

*Figure 9: Duration distributions as a function of the departure time (Smart ED, human health).*

*II.D. Modelling*

The generated profiles are the input for an implemented mobility model. The model simulates the driving behavior and the grid connection of EVs via the charging station. Figure 10 presents the general structure of the simulation:

- If a vehicle begins a trip due to the calculated probability, it gets random but correlated values of the day and time-dependent distributions of distance, duration, and the normalized consumption per kilometer. The new EV data are updated with the given trip values and the car is plugged to the charging station after the given trip duration.
- If a vehicle does not begin a trip, it charges in case the state-of-energy (SOE) is lower than the required energy for the mobility. This is explained in more detail in the following section II.E.

The charging process of the vehicle is simulated with real charging curves measured in our laboratory at PGS RWTH Aachen University. Figure 11 (left) depicts an 11 kW constant-power charge. The AC power on the grid side is larger than the DC power on the EV battery side, which shows the power losses due to the power converter. The efficiencies are around 93% for the constant power phase and decrease during the power decline. Battery current and voltage are shown in Figure 11 (right). The current values range in the case of this charge (11 kW) from 2 A to approximately 30 A at battery voltages of around 320 V to 390 V. During the constant power phase, current decreases while voltage increases with higher SOE.

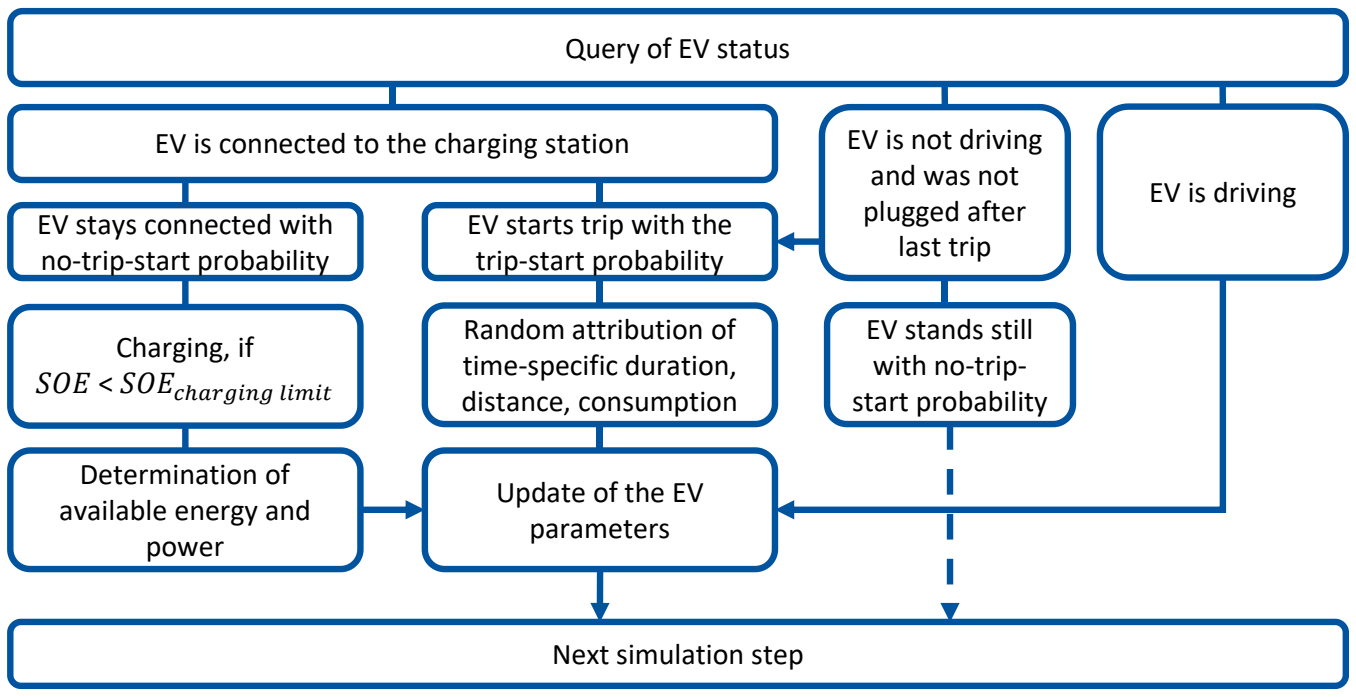

*Figure 10: Simulation flow of the implemented mobility model.*

It is important to consider that at around 90% SOE the constant-power charge turns into a constant-voltage phase resulting in a strong decline of current and power. When offering ancillary services, EVs with - in this case - SOEs above 90% are therefore not chosen to participate in the pool because of their strong decline in power. That is why a charging SOE limit of the EV is implemented in the simulation. This limit is set dynamically and ensures that all logged trips could have been done, individually for each EV. The charging limit is mostly around 80% of the SOE and is further explained in section II.E.

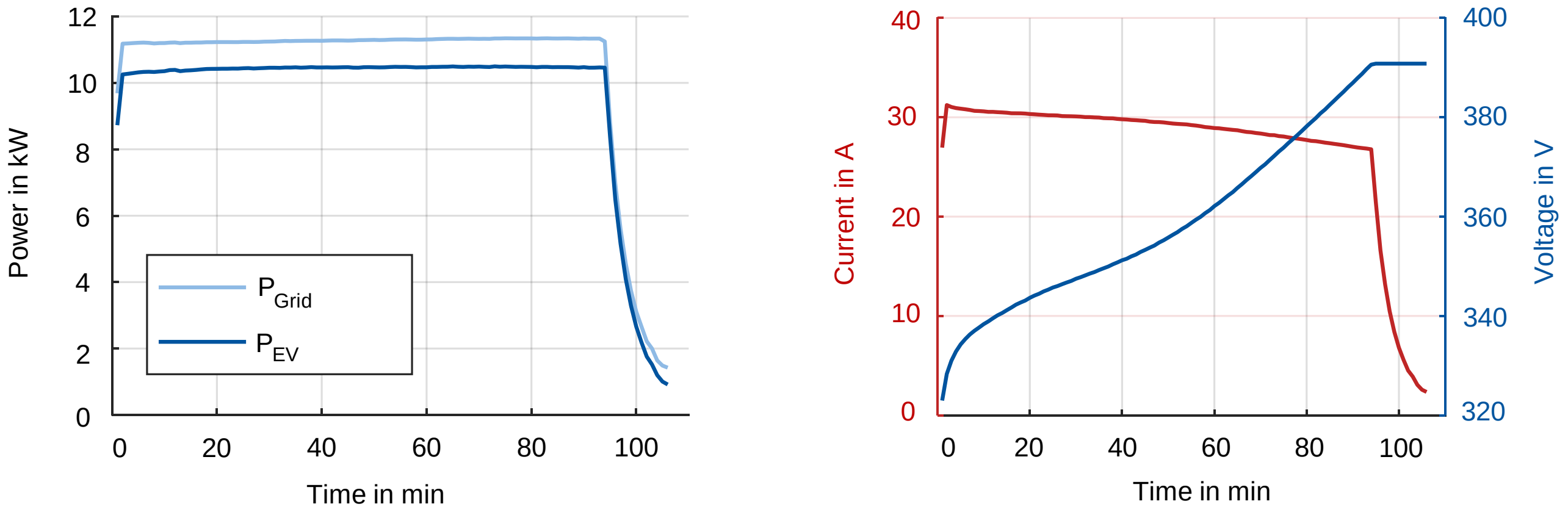

*Figure 11: Measured grid and battery power (left) and battery current and voltage (right) of a Smart ED during charge at 11 kW AC power.*

*II.E. Virtual sectioning of the battery*

The primary use of an EV is mobility. However, several studies as well as our presented data have shown that the average trip distance is quite low resulting in a low average energy needed for most trips. Thus, there are free energy capacities during most times that can be used for dual-use concepts in order to increase the economics of operation. Within these concepts, the primary use should not be limited by the secondary use. For dual use, the battery must be virtually divided into the energy range for primary use (mobility energy $E_{mobility}$) and the energy range for secondary use (marketable energy $E_{market}$), as shown in Figure 12. The mobility energy must be ensured at any time in order to undertake the regular trips of the vehicle. It consists of a reserved minimum energy for spontaneous trips at any time and a trip energy within certain time windows, when the vehicles are generally operated. The marketable energy, on the other hand, is the difference of the total battery energy $E_{Bat}$ and the current mobility energy as shown in equation (2). It is a derived quantity that describes how much of the battery's rated energy capacity a user would allow to use for a secondary use such as FCR.

$$E_{market}(t) = E_{Bat} - E_{mobility}(t) \qquad (2)$$

Based on the state of energy (SOE), the marketable energy can be divided into the charge energy $E_{charge}$ and the discharge energy $E_{discharge}$ as shown in equation (3)-(5) and in Figure 12 (left).

$$E_{market}(t) = E_{charge}(t) + E_{discharge}(t) \qquad (3)$$

$$E_{charge}(t) = E_{Bat} - SOE(t) \; \forall \; SOE(t) > E_{mobility}(t) \qquad (4)$$

$$E_{discharge}(t) = SOE(t) - E_{mobility}(t) \; \forall \; SOE(t) > E_{mobility}(t) \qquad (5)$$

In order to calculate the FCR power an EV can provide, some basic calculations are necessary that are described in the following. Generally, the FCR power can either be restricted by:

1. the rated power $P_{EV}$ of the battery converter of the EV,
2. the rated power $P_{CS}$ of the charging station, or
3. the charge or discharge energy in combination with the time of power supply.

Equations (6) and (7) consider all three cases for calculating the available charge power $P_{charge}$ and discharge power $P_{discharge}$ taking the service period $\Delta T_{supply}$ of the required 15 minutes full FCR power into account.

$$P_{charge}(t) = \min\left\{\frac{E_{charge}(t)}{\Delta T_{supply}}; P_{EV}; P_{CS}\right\} \qquad (6)$$

$$P_{discharge}(t) = \min\left\{\frac{E_{discharge}(t)}{\Delta T_{supply}}; P_{EV}; P_{CS}\right\} \qquad (7)$$

**Example: Assumptions for calculating the charge power of one time stamp:**

$E_{Bat} = 50$ kWh (before constant voltage phase starts)

$E_{mobility} = 15\ kWh\ (30\%\ of\ E_{Bat})$; SOE = 30 kWh

$P_{EV} = 11\ kW; P_{CS} = 22\ kW, \Delta T_{supply} = 15\ \mathrm{min} = 0.25\ h$

**Calculation:**

$E_{charge} = 50$ kWh – 30 kWh = 20 kWh

$$P_{charge} = \min\left\{\frac{20\ kWh}{0.25\ h}; 11\ kW; 22\ kW\right\}$$

$$= \min\{80\ kW; 11\ kW; 22\ kW\}$$

$$= 11\ kW\ (power\ limited\ by\ EV\ in\ this\ case)$$

For our simulation, we assume a bidirectional charging station. Although most of the current charging stations at the time this paper is submitted are unidirectional, broad literature expects EVs to play an important role in ancillary services of the future and bidirectional charging stations will probably emerge for V2G and vehicle-to-home (V2H) applications (see section I.B). However, many car manufacturers still do not provide any internal vehicle information such as the SOE to the charging station and prohibit the charge control as they use older Open Charge Point Protocols (OCPP version 1.5 or 1.6). Nevertheless, protocols like OCPP version 2.0 with the corresponding ISO 15118 protocol as well as the CHAdeMO protocol send information and allow charge and discharge control.

The required energy for mobility changes with the time of day and the day of the week depending on the use profile of the EV (see Figure 12). In order to take the individual use profile of the EV into account, the minimum energy required is calculated on the basis of historical journey data. In addition, a spontaneous mobility buffer of at least 30% of the battery energy is always implemented. This value is specified by users in a field test as the desired minimum for spontaneous trips [30]. Whenever the SOE is above the mobility energy, the vehicle can provide grid services as long as the SOE is not within the $E_{CV}$, where the constant-voltage charging phase takes place (see Figure 11). In order to provide flexibility in charging and discharging the EV, an additional individual charge limit of about 80% SOE is introduced (see Figure 12 (right)). This upper limit is chosen as high enough that sufficient energy for all historical journeys is available. The individual charging limit ensures that the single EV is not yet in the constant-voltage phase of the charging process.

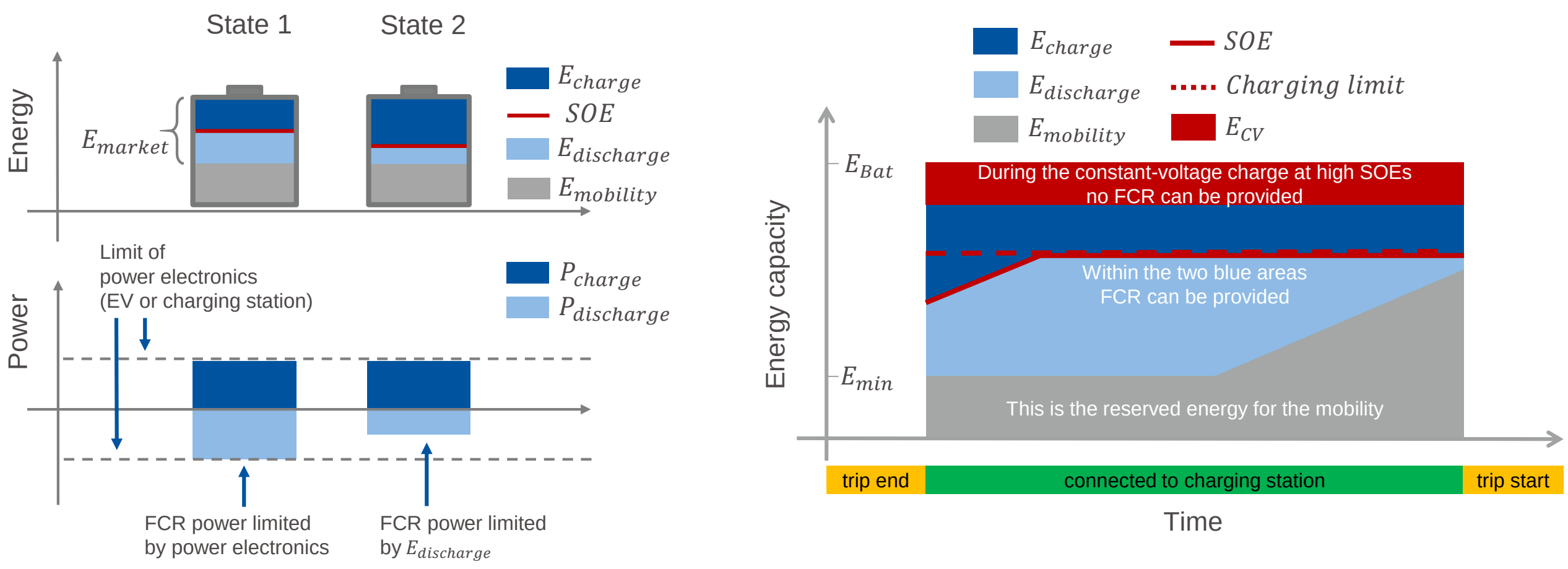

*Figure 12: Virtual division of the battery energy capacity and calculation of available power. Left: static. Right: time-dependent in the case of EVs. Inspired by [30].*

## *II.F. Frequency containment reserve*

This section presents the German FCR market design, its requirements for BSSs and the development of FCR prices.

### *II.F.1. The German FCR market*

The frequency regulation is divided into frequency containment reserve (FCR, former primary control reserve), automatic frequency restoration reserve (aFRR, former secondary control reserve), and manual frequency restoration reserve (mFRR, former tertiary control reserve). The three types of frequency regulation have different activation times and replace each other consecutively as shown in Figure 13. Within 30 seconds after a frequency deviation of more than 10 mHz, FCR units in Continental Europe Synchronous Area have to provide FCR automatically [57,58]. This way the frequency drop (respectively rise) is supposed to be stopped. Providers of FCR must offer both positive and negative FCR power for the same service period. It is important to notice that other regions such as the Nordic Balancing Markets [59] or the UK [60] also have faster frequency regulation markets.

The regulatory requirements for FCR varied during the last years. Appendix, Table 11, provides the decisions taken on FCR by the German Federal Network Agency (FNA) and the TSOs. Until mid-2011, FCR was tendered on a monthly basis in a pay-as-bid auction, which means that the supplier of FCR had to provide the service for one month continuously and the paid prices were the individual prices that providers bid in the auction [61,62]. From mid-2011 until July 2019, FCR was tendered weekly in a pay-as-bid auction [61,63] and the minimum bid size was decreased from 5 MW to 1 MW [61].

In July 2019, the service period was shortened to one day [61] and the pricing was modified to a market-clearing-price procedure for the offered power [63]. This means that every provider of FCR earns the price of the highest offer that is accepted for the respective bidding period. As a last modification, in July 2020, the service period was made even more flexible to six daily slots of four hours each [63]. The EU had demanded this higher flexibility and short-term nature of FCR tenders [64].

### *II.F.2. Requirements on BSS and virtual power plants when participating in the FCR market*

BSS are technically able to provide frequency regulation due to their fast reaction times and high cycle stability, which is required when, for example, offering FCR [9,65,66]. In 2015, the German TSOs had decided on a 30-minute-criterion for BSS, when providing FCR [67]. As the storages had to provide the positive and negative power simultaneously over the period of one week, the TSOs wanted the BSSs (or pools of BSSs) to be able to provide the awarded power for at least 30 minutes (instead of the usual 15 minutes) [67]. Only when BSSs were added to an existing pool to increase its flexibility, 15 minutes were sufficient [67]. In 2019, the German FNA rejected this request of the TSOs as it would have discriminated the BSSs operators [11]. In addition to this minimum required amount of energy, an FCR provider must also maintain an additional quarter of its prequalified power as a buffer [68]. This is to ensure that storage management activities can be provided at the same time as FCR provision. For example, a pool of EVs that wants to provide 3 MW of FCR power must have at least (3 MW · 1.25 =) 3.75 MW available. For a pool, the TSOs set a minimum size of 25 kW for the smallest plants and 2 MW for the pool [69]. A system may only participate in one pool at a time [69].

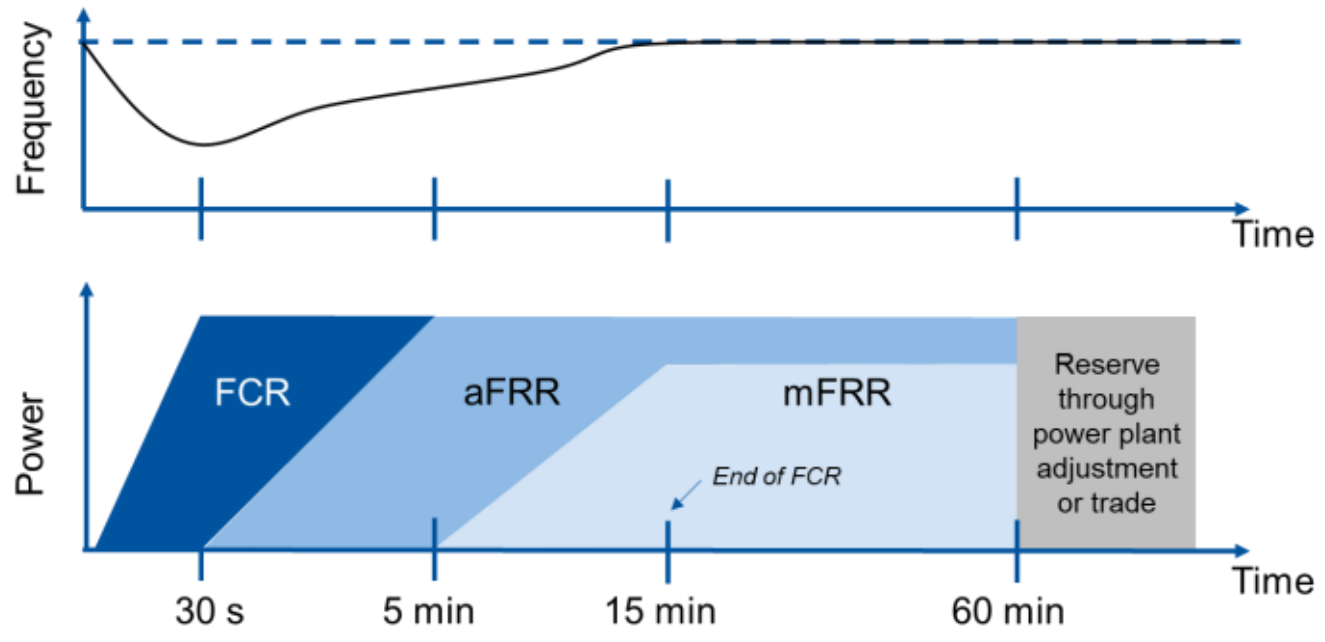

*Figure 13:Division of frequency regulation with exemplary frequency curve (top) and power type responsibilities (bottom) based on [57]. Figure shows only a frequency drop, although FCR is bidirectional.*

*Table 4: Frequency Containment Reserve Market before July 2019, after July 2019 and after July 2020 [61,63]*

| | Before July 2019 | July 2019 – July 2020 | Since July 2020 |
|---|---|---|---|
| Direction | Positive and negative power together | | |
| Minimal bid | 1 MW | | |
| Minimal increment | 1 MW | | |
| Reaction time | 30 seconds | | |
| Provision time | 15 minutes | | |
| Remuneration | **Pay-as-bid for power** | **Market-clearing-price for power** | |
| Time sectioning | **1 week** | **24 hours** | **4 hours** |
| Tendering | **Tuesdays, 3pm** | **D-2, 3pm,** | **D-1, 8am[1]** |
| Demand Germany | 551 MW - 620 MW | 620 MW | 573 MW |

*II.F.3. Development of the prices for FCR*

Figure 14 (left) shows the development of the FCR prices from 2008 to 2020. It contains all special features of the market history: the different service periods are marked by different colors. In addition, the range of values of 95% of all prices shows that the monthly and weekly service periods still had a pay-as-bid price in contrast to the market-clearing price of the daily and four-hourly service periods. The FCR prices show volatility over the shown time period. Nevertheless, a clear trend towards falling prices can be seen since 2015. While prices in 2015 still averaged around 3,600 €/MW/week (peaks above 6,000 €/MW/week), by 2020 they had fallen to less than 1,500 €/MW/week. This is mainly due to the strong competition in the FCR market caused by the increasing number of large-scale BSSs [2].

Figure 14 (right) shows the prices for different time spans of the recent three service periods:

1. Weeks (2018/07-2019/06)
2. Days (2019/07-2020/06)
3. Four hours (2020/07-2020/12)

The prices in the last year of weekly tendering were significantly higher than for other periods. While the prices averaged around 1,880 €/MW/week during weekly service periods, the average price during daily service periods was around 1,280 €/MW/week and during the service period of four hours 1,370 €/MW/week. Further analyses show that the prices of the service periods of four hours are around 30 €/MW/4 h for all six daily time slots. Thus, the current market design shows that the different four-hour time slots do not cause major differences in prices. This is probably due to the fact that a stationary large-scale BSS must bid at all times when the FCR is its only revenue stream. However, this characteristic could be changed in the future, especially by many small, decentralized storage systems like home storage systems or EV pools (see section II.F.2). In the near future, especially during night hours, there will be a large number of small storage units in the grids that could provide FCR. Since this is only an additional revenue stream to their primary use, we expect prices to drop significantly at these times, as the small units do not rely on FCR revenue to refinance the bulk of investment costs, but only variable costs for providing FCR.

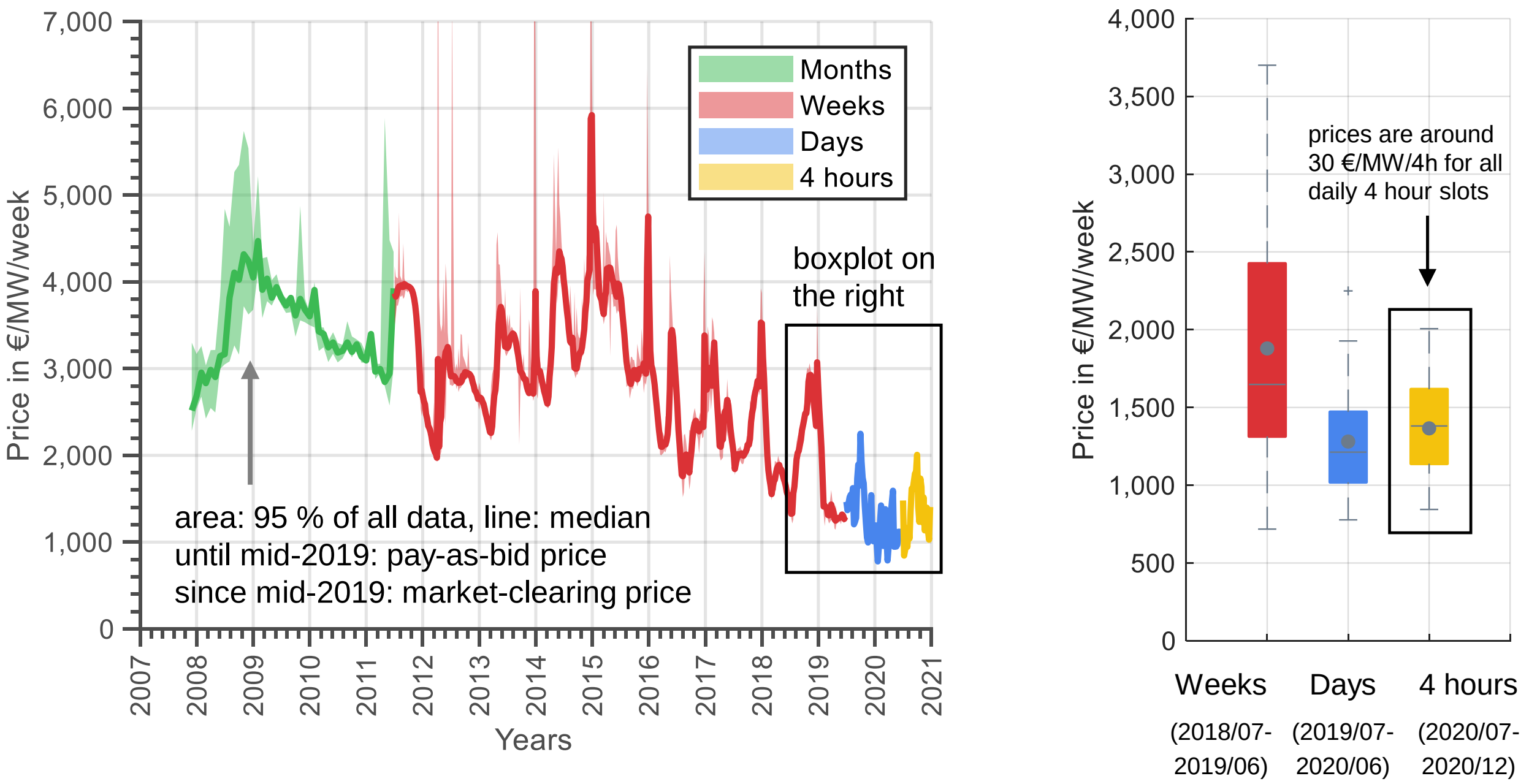

*Figure 14: Left: FCR price development per week. Analyzed data from [7]. The prices of the periods of one month, one day and four hours are scaled to a weekly price for comparability. Therefore, the monthly prices are divided by four, and the daily and four-hourly prices are summed up within the respective week.*

*Right: FCR prices for a service period of four hours from 01. July 2020 until 31. December 2020. Analyzed data from [7].*

## III. Results

In this section the results are presented and discussed. First, the influence of flexibilization on the available power of an EV pool is examined. Subsequently, these results are used to calculate the revenue using the example of German FCR prices. These prices are representative for many countries in Central Europe.

### III.A. Available power

As the EVs are on trips for several parts of the day, the available pool power fluctuates. Figure 15 shows the distributions of the minimum bidirectional pool power for four exemplary clusters. Therefore, this power is the minimum of both the positive and negative power the pool could provide. In the following, the minimum bidirectional pool power is called *“power capability profile”.* The results shown are based on an annual simulation with the simulation interval of 15 min, which corresponds to the 15-minutes criterion for FCR (see section II.F). The distribution shows all days of the year over a period of 24 hours. While the median is represented by the thick line, the differently colored areas show the respective ranges of values from the legend of 50%, 75%, and 100% of all values. In the corresponding figures, the clusters manufacturing, human care, public administration, and other services are shown, since these contain the highest number of vehicles and are therefore most representative. All vehicles have in common that the vehicles are more connected to the charging stations overnight than during the day. Nevertheless, the available power at night tends to be around 80%, as the vehicles are not always parked on company grounds. The profiles manufacturing, public administration, and other services are relatively similar. Overnight, the median power is between 80% and 90%, and during the day, due to the working hours, it is between 40% and 50%. The confidence intervals cover a wide range of values from about 40% to 90%. Especially during the day, there are large ranges of values, as some vehicles are seldom or not driven at all during the weekends, for example. However, the profile of the human health cluster is clearly different. First, the median shows two shifts (morning and afternoon). In addition, the regularity of the trips is well illustrated by the narrow confidence intervals. The human health service also has to care for patients on weekends, which means that its profile is the same seven days a week. During the morning shift, minimum values of less than 20% of the cluster power occur, while maximum values between 60% and 80% of the cluster power occur at night. During the afternoon shift, the median is around 55% of the power.

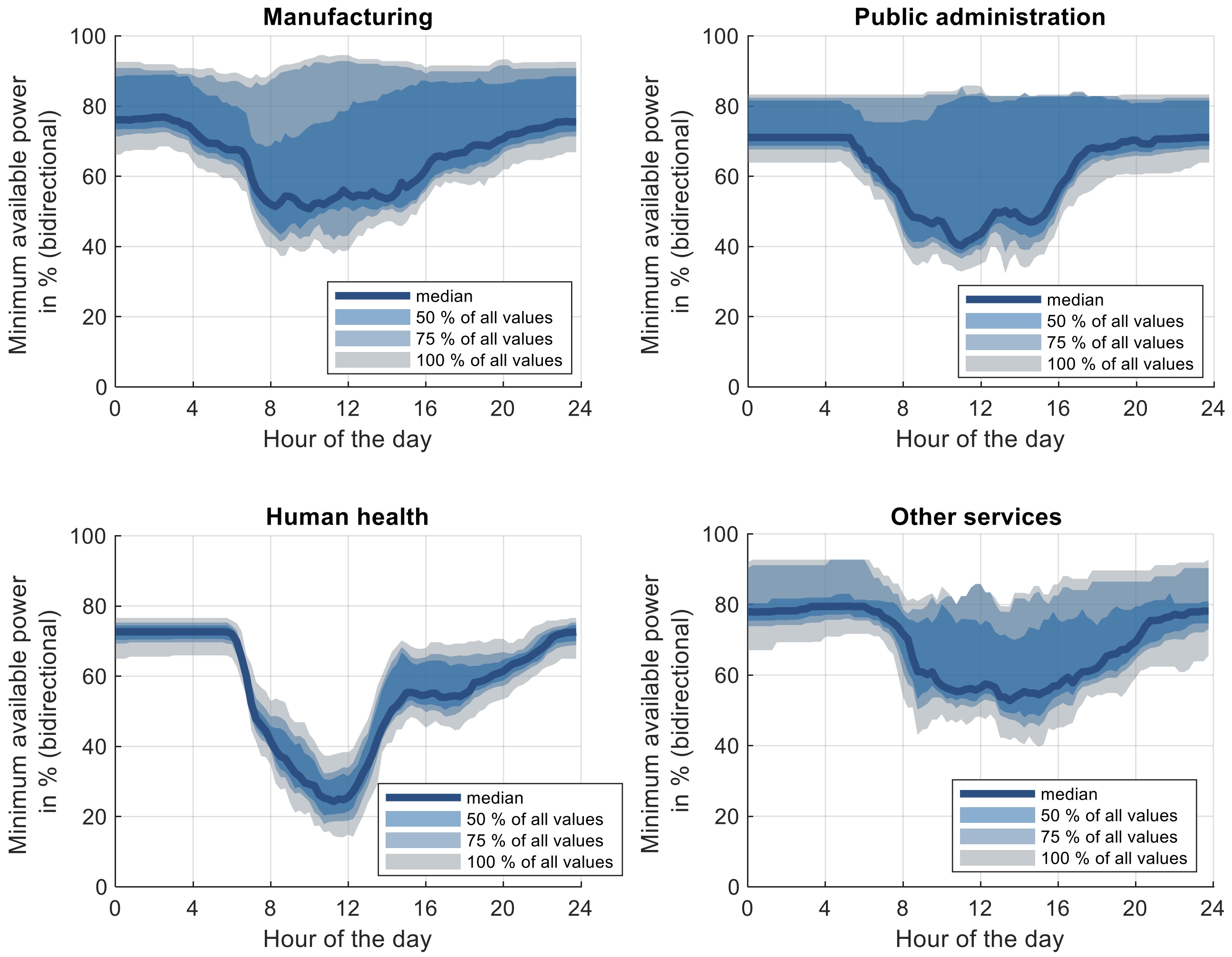

*Figure 15: Power capability profiles of the four clusters manufacturing (top left), public administration (top right), human health (bottom left), and other services (bottom right) for all weekdays.*

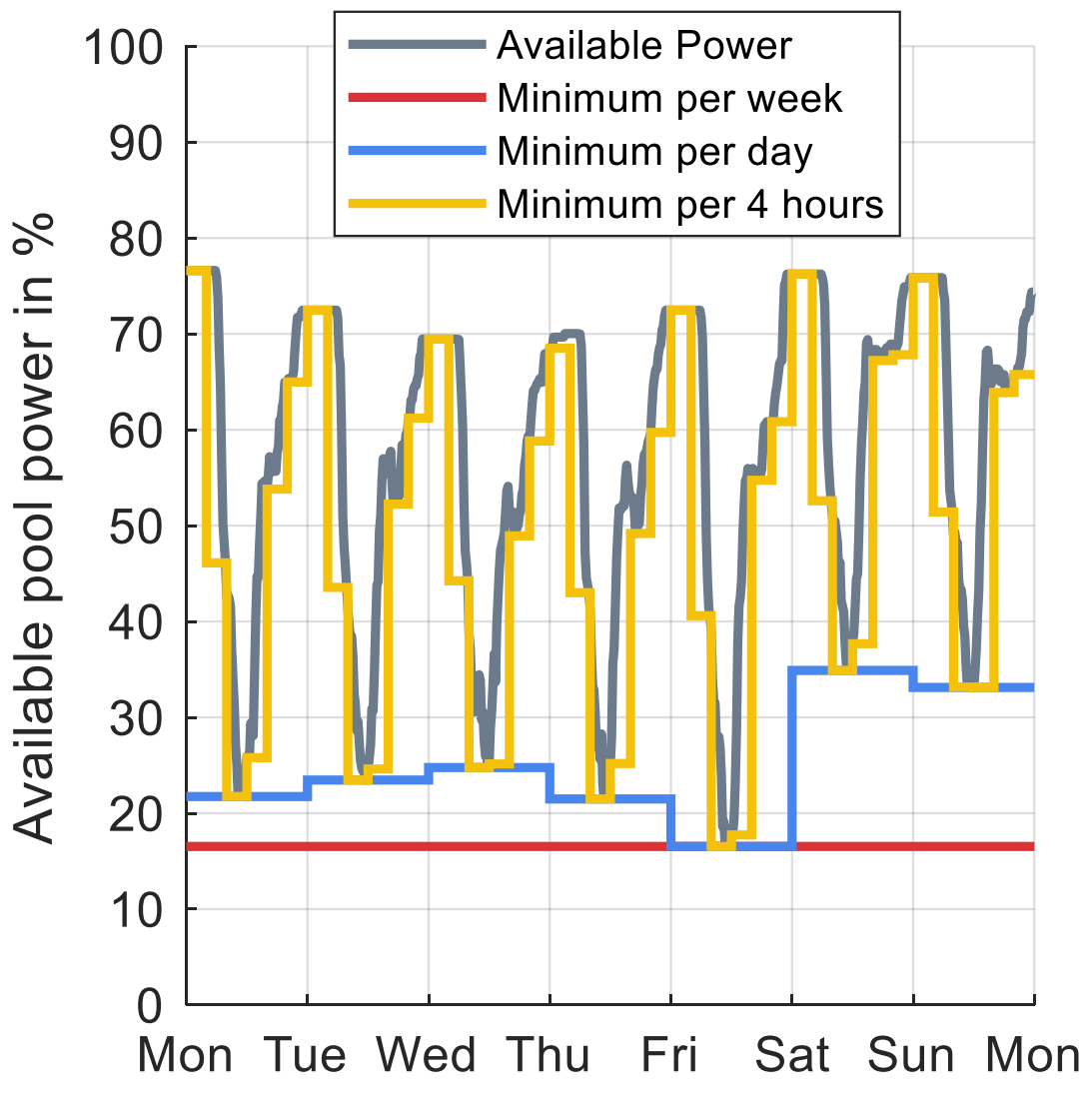

*Figure 16: Available pool power for different service periods for human health service cluster.*

The four profiles already show that EVs are particularly suitable for short periods of ancillary services, as the availability of power varies with the time of day. The median can be seen as the power to be expected and the size of the confidence interval as the certainty with which an offered service can be provided (a large confidence interval corresponds with large uncertainty).

Based on the time series of the power capability profiles, the influence of FCR flexibilization can be further investigated. The minimum power within a service period represents the available FCR power. Figure 16 illustrates the impact of shorter service periods using the human health power capability profile for an exemplary week. While the absolute minimum of the week determines the FCR power to be marketed in the case of a service period of one week, shorter service periods allow the fleet to be used even at times when many vehicles are connected to the charging stations and can provide power. The power minimum of the weekly service period is less than 20% of the pool power. Since the EVs in this case drive about the same seven days a week, the flexibilization from weekly to daily service periods provides on five days (Monday through Friday) only a slight increase in the daily minimum to about 25% of power. On weekends, slightly fewer trips by the pool EVs provide a minimum above 30%. The FCR power at the service period of four hours can follow the volatile profile much better and even corresponds to the available pool power of 70% to 80% at night. At these times, the available power does not correspond to 100%, since the vehicles are often not plugged overnight.

Figure 17 aggregates these evaluations for a whole year and shows the ranges of available power within the year. The influence of the service period on the available power is shown for four different clusters. In total, there are four “real” service periods of the historical and current FCR market designs (1 month, 1 week, 1 day, 4 hours) and two fictitious shorter service periods of 1 hour and 15 minutes respectively. The two performance periods of one month and one week are very similar and the available powers are close to the absolute minimum. This is because such long periods often result in a situation where many vehicles are on the road at some point. For this reason, the respective clusters can only offer little power to be able to guarantee service even under worst-case conditions. In such cases, a pool would have to be significantly oversized. The change from weekly to daily service periods leads for the clusters manufacturing, public administration, and other services already to significantly more time slices at which high powers can be offered. These are especially the weekends, as further analyses show, where the cars do not drive that often. The change from weekly to daily service periods has significantly less effect on the available power of the human healthcare service. This is because it operates nearly the same seven days a week and a service period of one day does therefore not bring much improvement. However, if the service period is further reduced to four hours, the power that can be offered by the human healthcare service also increases significantly. This increase is particularly due to the night when most of the EVs are connected to a charging station. Since the idle times are usually longer than eight hours, the further flexibilization through the 1-hour and 15-minute delivery periods will only result in slight increases in the available power. The increase in power thus shows a decreasing sensitivity to the shortening of the performance period.

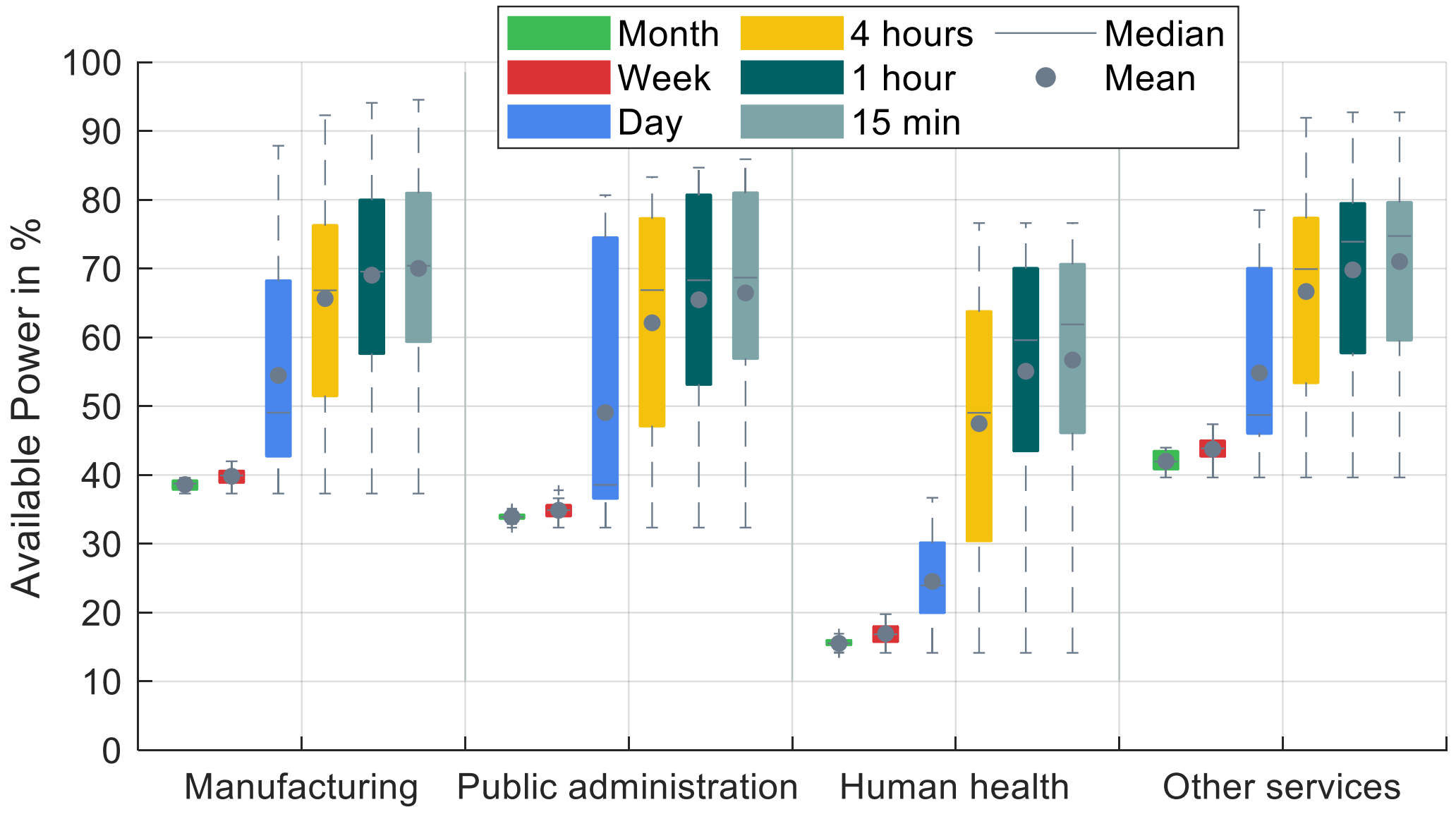

*Figure 17: Available pool power of the EV operating in the chosen sectors.*

### *III.B. Achievable revenue*

The FCR prices have been decreasing for years (see Figure 14) and the available pool power increases with shortened service periods (see Figure 17). Therefore, this section examines the development of the theoretical revenue of a 1,000 EV pool with a charging power of 11 kW per EV to analyze the counteracting forces of decreasing prices and increasing power availability from flexibilization. Three service periods (week, day, and four hours) from mid-2018 to the end of 2020 are analyzed in detail.

Figure 18 shows an example of the power capability profile for the human health service and the FCR price for a day in 2020 with a service period of four hours. The following steps are taken to calculate the revenues. It becomes obvious that the FCR potential of the pool is limited both by the required buffer power (step 2) and the increment condition (step 3).

1. Determination of the minimum power in the respective service period.
2. Division of the minimum power by factor 1.25 in order to ensure the required 25% power buffer.
3. Rounding down the power rest to an integer multiple of 1 MW to correspond to the minimum power of 1 MW and the increment of 1 MW in the current market design.
4. Multiplying the time-dependent FCR power by the time-dependent FCR price to determine revenues.
5. Summation of the revenues for the entire period and normalization to one week.

This procedure is performed for the whole time series of the annual simulations (in the case of the service period of four hours the simulation is half a year) of all clusters for the three selected service periods of one week, one day and four hours with historical prices.

Figure 19 shows the results of the revenue per pool for the three service periods studied. The plot shows the achievable revenue per EV and year for all clusters based on an annual simulation of a 1,000 EV pool for each cluster. Each box plot contains the 14 clusters for the different service periods. It can be seen that the increasing flexibility of the examined clusters overcompensates the lower prices and that revenues increase on average. The average revenue per EV increases from about 180 €/a with weekly service periods, over 232 €/a (daily service period) up to 346 €/a with four-hour service periods. Overall, the revenues are between 27 €/a and 405 €/a per EV. Furthermore, the overall influence of flexibilization and falling prices clearly depends on the individual service profile as can be examined even more clearly in Figure 20:

- The change from weekly to daily service periods was accompanied by increasing flexibility and falling FCR prices. For the majority of the clusters the increased flexibility overcompensates the falling prices with mean revenue increases of 52 €/a and maximum increases of around 177 €/a per EV. However, there are also some clusters for which the falling prices dominated and their pool revenue decreased by approximately 41 €/a. These are clusters with a relative constant power capability profile that cannot provide much more power if the service periods get shorter as they do not have volatile power peaks.
- The change from daily to four-hourly service periods was accompanied by increasing flexibility and slightly increasing FCR prices probably due to the new market design. Both developments lead to an increase in revenues from 85 €/a to 167 €/a (mean: 114 €/a) per EV. In this case, volatile profiles that differ from day to day could benefit both from offering higher FCR power as well as from increasing prices. However, profiles that are either quite constant over a week or that have the same volatile profile day by day did only benefit from the price increase. While the first ones do not need shorter service periods at all, the latter ones need even shorter service periods to offer their power peaks that occur within a day.
- The overall change from weekly to four-hourly service periods was accompanied by increasing flexibility and falling FCR prices. The flexibility overcompensates for the falling prices with mean revenue increases per EV of around 166 €/a.

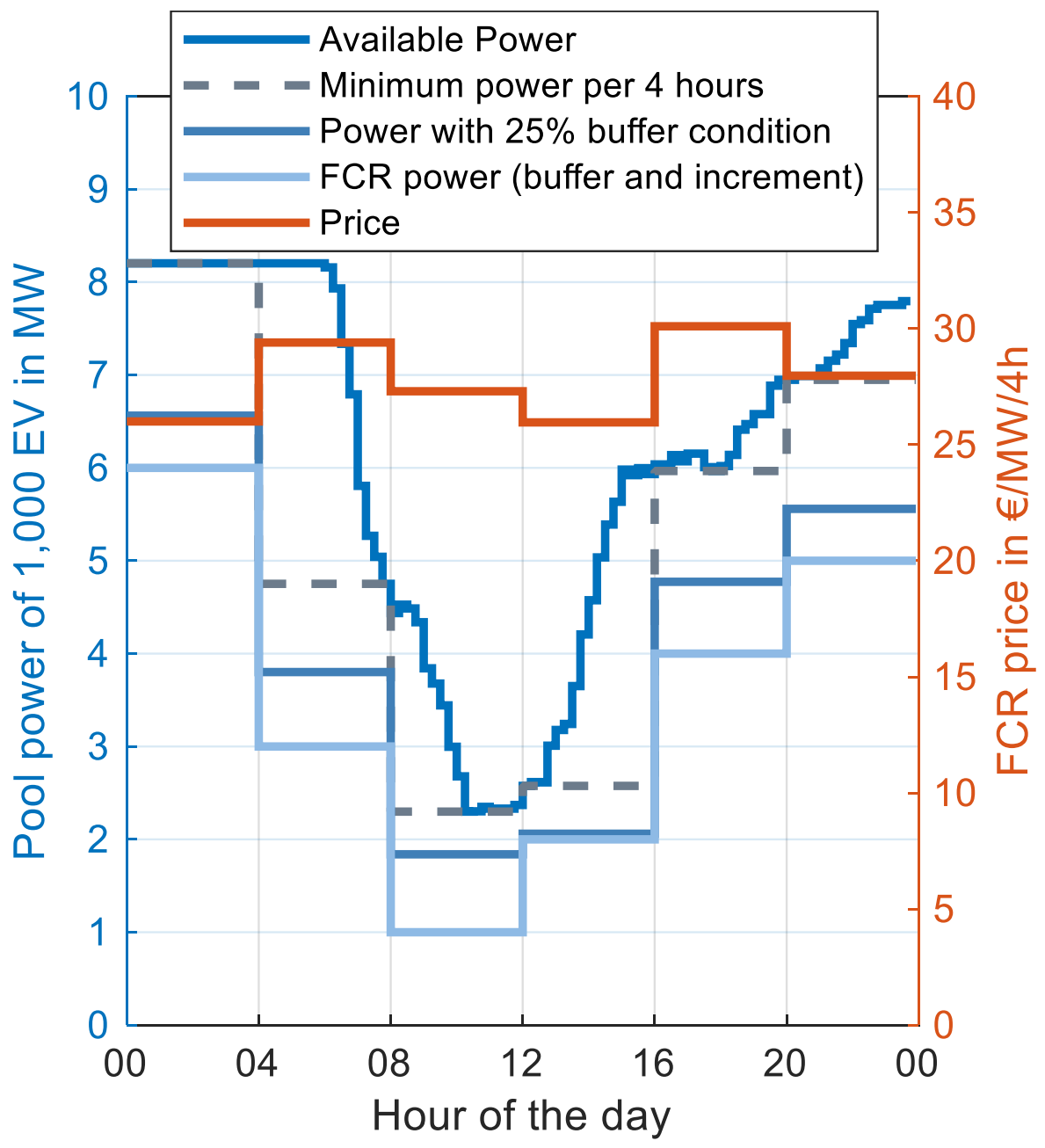

*Figure 18: Available pool power for different service periods (human health).*

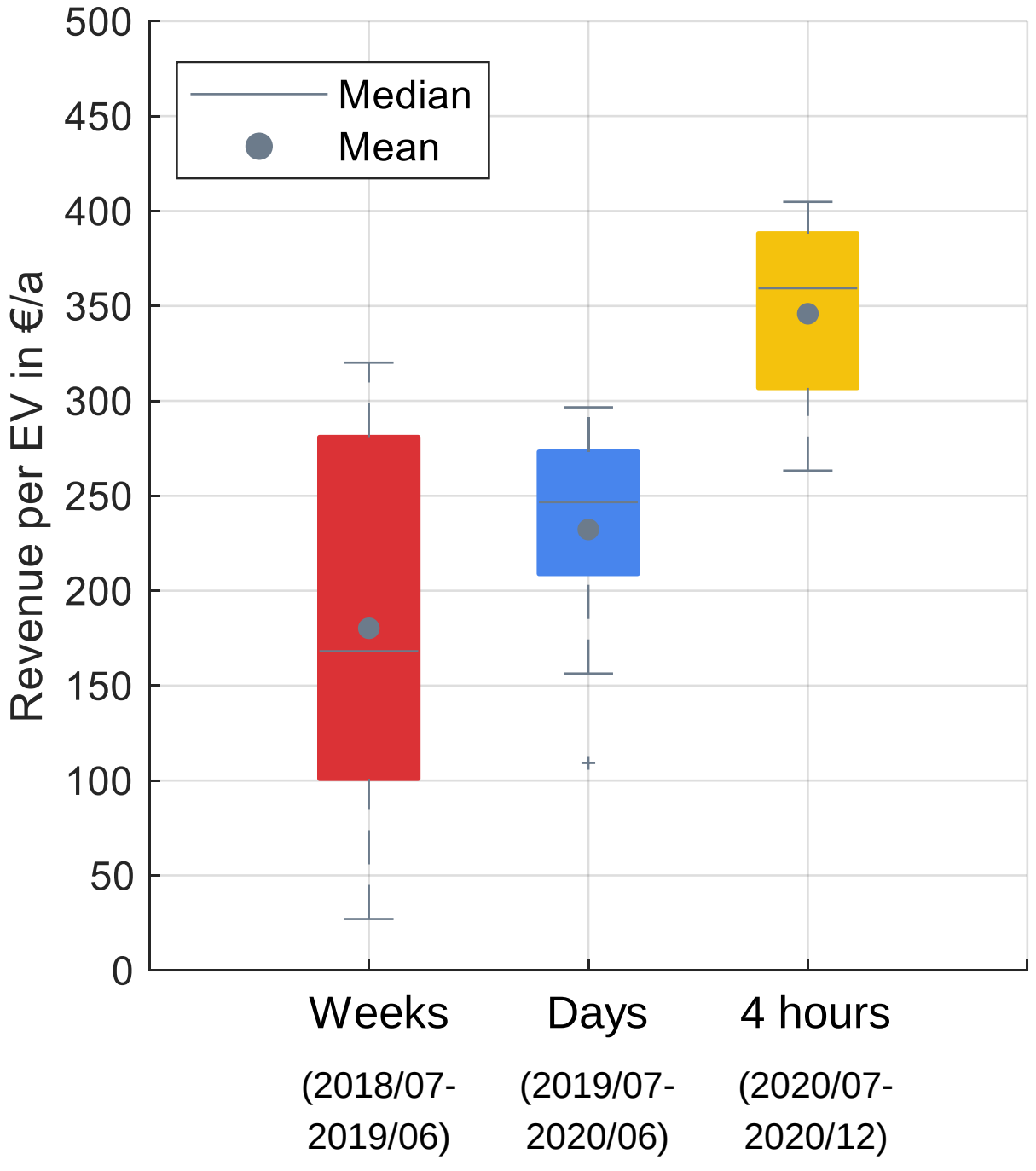

*Figure 19: Yearly revenue per EV based on simulation of 1,000 EVs over a whole year. Each box plot contains the 14 clusters.*

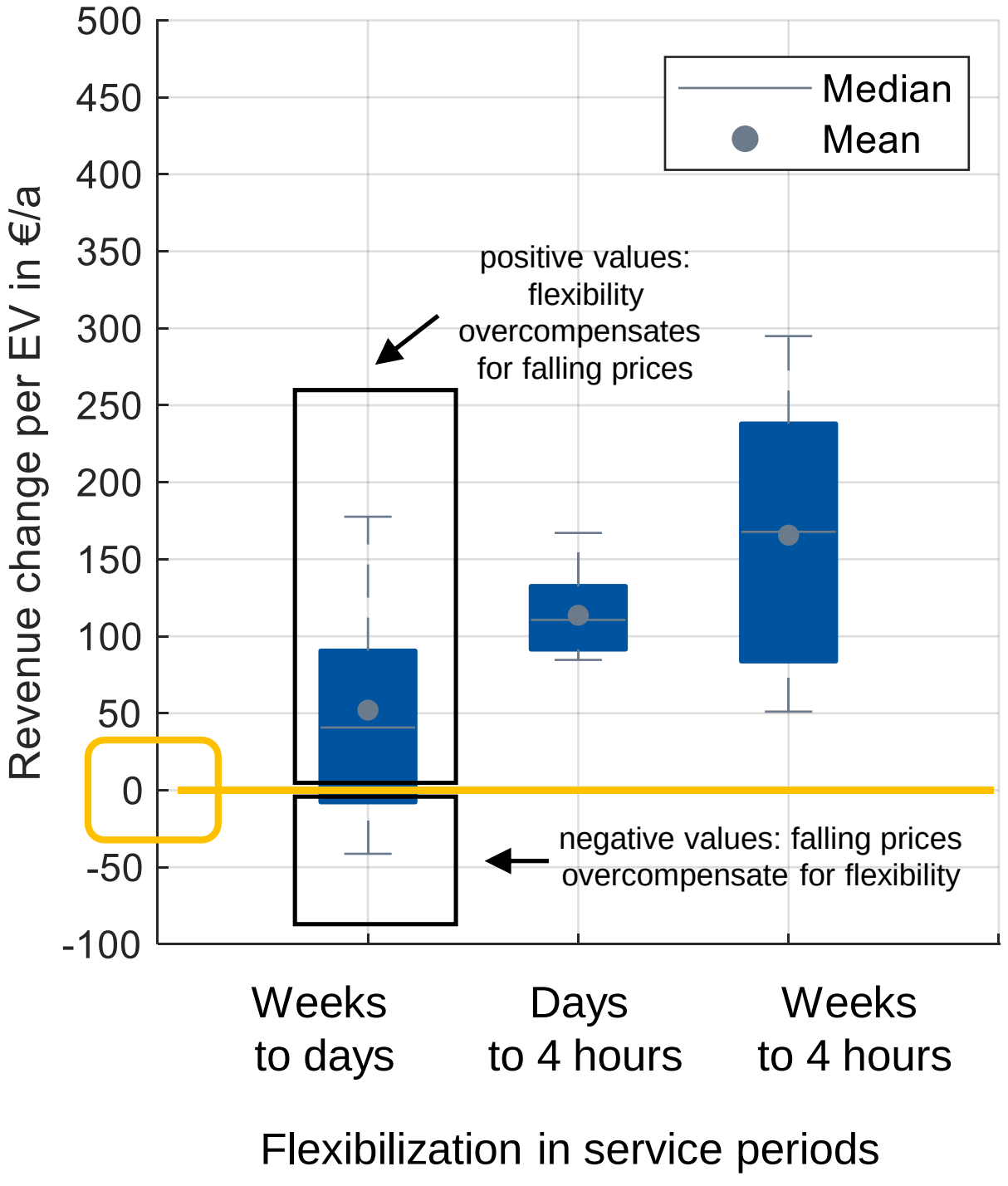

*Figure 20: Change in weekly revenues per EV based on simulation of 1,000 EV pool over a whole year. Each box plot contains the 14 clusters.*

Especially the revenue for pool profiles that have a volatile profile within a day such as the human healthcare increased by up to 295 €/a per EV. These profiles can now use, for example, the night hours during which idle times are longer than the service periods of four hours (see Figure 16).

This illustrates that a general trend of dominating flexibility can be recorded. However, it depends on the concrete pool profile: In the pools where EVs are less frequently driven the opposing movements of falling prices and increasing flexibility roughly balance each other out. Volatile profiles such as the one of the human health pool, on the other hand, show almost a doubling of revenue from the weekly service period to the four-hourly period as their idle times are longer then the short service periods of four hours. This can be seen in Figure 16, as the EVs have the same profile 7 days a week and run two shifts a day. Analogous to the available power in Figure 17, there is only a small change in revenue when changing from weekly to daily service periods. However, the four-hour delivery period allows the frequently driven EVs to regularly use the hours of night standstill to market FCR. Thereby the economically positive effects of the flexibilization exceed the falling prices clearly.

## IV. Discussion

In this paper, we estimated the revenue potential of EV fleets using historic market prices, mobility profiles of different fleets, and EV characteristics such as the battery size, charging power, and consumption.

However, a few points need to be considered for further classification of the results. These relate to techno-economic points and aspects of the charging strategies.

### IV.A. *Techno-economic aspects*

In our study, we focused on achievable revenues. These revenues are necessarily offset by associated costs. The costs are mainly composed of hardware and transaction.

For the provision of FCR, a bidirectional charger for the EV is needed. Although, as of 2020, most EVs are not used for bidirectional activities yet, but this situation could change quickly. Volkswagen, for example, announced that their EVs will have bidirectional features from 2022 onwards [70]. First bidirectional products for private car owners supporting both standards (CHAdeMO or ISO 15118) already exist [71] (around 6,000 € per 7,4 kW wallbox [72]) and they will most likely become cheaper with a growing market.

Besides hardware, there are also operational expenditures such as metering costs, costs for pool management by an aggregator, and battery degradation due to increased battery cycling through FCR (around 250 equivalent full cycles per year for large-scale BSSs [1,66]). The battery degradation costs of two field projects are estimated to be 50 €/a – 100 €/a in [30] and 86 €/a in [73]. Further, the degradation is mainly influenced by calendar aging and not by cycle ageing through V2G [73]. The only known source to the authors dealing with operational costs for EV dual use is the INEES project [30], in which the provision of aFRR through EVs was analyzed. The cost estimation for metering, communication, and battery degradation summed up to 700 € to 750 € per year for 2016 and is projected to be 110 € to 350 € per year for a future scenario with bidirectional EV and charging stations. Estimating possible best-case revenues per EV for the analyzed commercial fleets from Figure 19 at maximum 405 €/a, the difference for the operational costs as of 2020 could probably be around some 100 € per year. This estimate fits well to reported real-world income of 21 € per month (252 €/a) in 2019 for the most earning private EV providing balancing power in the Netherlands for the company Jedlix BV [74]. All in all, the profits can possibly be in low positive ranges, if capital expenditures are neglected. However, this best-case scenario implies that the aggregator is always awarded a contract in the FCR tendering, which is rather unlikely with the growing number of batteries in the market, especially in the future. Besides these costs, high penalties and exclusion of the FCR market could occur, if FCR power cannot be provided in case an unforeseen high number of EVs is on the road. Aggregators should therefore have either absolutely plannable vehicle fleets such as busses or more reliable assets such as stationary BSSs or other power plants in their portfolio.

### *IV.B. Charging strategies and "degrees of freedom"*

In our estimation we assumed that the pool operator does not plan an optimal charge management of each individual EV. Therefore, the estimated revenue is the potential for a fleet with an undisturbed charging profile. The active charging management of the EVs by the pool operator offers the possibility to increase the power available for FCR provision which in turn would increase the revenue. In [10] such a real-world operating strategy of FCR provision for a large-scale BSS is presented and discussed for the historic market design with the 30-min criterion. With an active charging management of each individual EV a pool operator could keep the fleet in a valid operating range to fulfill the 15-minutes criterion more often and to increase the pool's FCR power. Charge management often requires energy trading, e.g., from the continuous intraday market. One study [75] describing large-scale BSS operation in FCR market shows results where expenses for intraday recharge, trading services, and connection to trader sum up to about 15 % of income when operating 4 MW/4 MWh storage capacity.

Another way of charge management is the use of the "degrees of freedom" in the provision of FCR [76]. The ENTSO-E Handbook requests a minimum accuracy of the frequency measurement of 10 mHz. Therefore, FCR does not have to be provided if the deviation of the frequency is within 10 mHz from the nominal frequency of 50 Hz. However, FCR can be provided within this so called deadband. With the use of an accurate frequency measurement the charge management could opt to charge EVs with FCR in the deadband which reduces the costs of EV charging for users and can be regarded as additional revenue. Furthermore, due to power measurement accuracy limitations, an overfulfillment of provided FCR power of up to 20 % is permitted. Also, this degree of freedom could be used to maximize the energy gained for EVs during the provision of FCR and be seen as additional revenue. The last degree of freedom that can be taken advantage of by a pool operator is the specified ramp rate due to regulations. In the case of FCR, a total activation of the required power has to be activated within 30 s. A highly flexible unit, such as the battery of an EV, can react instantaneously in order to maximize energy when FCR is used to charge the EV. As an example for the impact of the degrees of freedom, one study [10] found that for a provision of 4 MW in the year 2014 with a large stationary storage system in Germany the energy gain due to the use of the degrees of freedom was 139 MWh. This rather complex topic for a real-world operating management for the provision of FCR with an EV fleet is a worthy research topic for the future.

## V. Conclusion and Outlook

This section draws a conclusion of the presented analyses and gives a brief outlook on market developments and future works.

### *V.A. Conclusion*

Traditional grid services are undergoing a change of auction design towards flexibilization. A few years ago, the market for frequency containment reserve (FCR) to stabilize the grid frequency in Germany was provided exclusively by conventional power plants over service periods of up to one month. At present time, many large-scale battery storage systems as well as some battery pools are participating in the same market with service periods of less than one day. The market for FCR was a promising source of income for battery storage a few years ago. However, prices have fallen significantly in the face of the sharp increase in competition from battery storage systems, and the sole refinancing of battery storage systems to provide FCR is becoming increasingly difficult. This paper investigated the influence of FCR market flexibilization and decreasing FCR prices on the economics of EV fleet operation.

The service periods were shortened from one week over days to four hours in accordance with the flexibility levels already achieved in the years 2018 to 2020. The average FCR price fell from 1,880 €/MW/week during weekly service periods to an average price during daily service periods of 1,280 €/MW/week to an average price of 1,370 €/MW/week during the service period of four hours.

The flexibilization from one week to one day causes an increase in available power of about 30% to 40% especially in the investigated profiles that have different driving profiles during the week. However, for EV fleets that have the same driving pattern seven days a week, the further FCR flexibilization to service periods of four hours is needed to significantly increase available power. The increase in flexibility from one day to four hours leads to a doubling of the available power for the examined human healthcare cluster profile, which drives two shifts each day a week, but is often connected to charging stations at night.

In general, flexibilization overcompensates for falling FCR prices and leads to higher revenues. While the potential revenue was on average in the range of 180 € per EV per year during the weekly service periods from mid-2018 to mid-2019, the mean revenue per EV per year increased to 346 € for the service periods of four hours from mid-2020 to end-2020. However, in all analyzed scenarios, the revenues are relatively low, and it remains unclear if they can overcompensate for the costs for metering, battery degradation, and pool management. At the moment, FCR provision through commercial EVs does not seem like a viable business model despite increased market flexibilization.

*V.B. Outlook*

In the future, we expect the flexibility of the spot and ancillary service markets to increase further. In parallel with a rapidly increasing number of EVs, there will be a huge potential of mobile BSS in the energy system for the near future. From an economic point of view, it is advantageous to use EVs during their idle times for grid service instead of leaving this potential unused. With respect to these developments, it is questionable which flexibility markets will remain and in what form and whether there will be new markets. The analysed FCR market, for instance, has a volume of only 600 MW. The regulatory agencies could decide to demand from EVs that they should have a frequency-dependent charging power profile. Such a law would effectively eliminate the FCR market as it is today. Such regulations are already part of the German renewable energy law (EEG), for example, which limits the feed-in power of photovoltaic systems. Furthermore, competition for aggregators will increase significantly. If there is enough battery capacity in the energy system, efficient pool management is essential. Therefore, in our future work we will investigate the influence of an optimal pool composition and management in different markets.

## CRediT authorship contribution statement

**Jan Figgener:** Conceptualization, Methodology, Software, Formal analysis, Investigation, Data Curation, Writing – Original Draft, Visualization. **Benedikt Tepe:** Conceptualization, Methodology, Software, Formal analysis, Investigation, Data Curation, Writing – Original Draft, Visualization. **Fabian Rücker:** Conceptualization, Methodology, Data Curation, Writing - Review & Editing. **Ilka Schoeneberger:** Conceptualization, Methodology, Data Curation, Writing - Review & Editing. **Christopher Hecht:** Methodology, Writing - Review & Editing. **Andreas Jossen:** Resources, Writing - Review & Editing, Supervision. **Dirk Uwe Sauer:** Conceptualization, Methodology, Resources, Writing - Review & Editing, Supervision.

## Declaration of Competing Interest

The authors declare that they have no known competing financial interests or personal relationships that could have appeared to influence the work reported in this paper.

## Acknowledgement

This work was partly done based on work within the research project "GO-ELK" (funding number 16SBS001C) funded by the German Federal Ministry of Transport and Digital Infrastructure and partly done within the research project "ALigN" (funding number 01MZ18006G) funded by the German Federal Ministry for Economic Affairs and Energy (BMWi). The authors of this publication are solely responsible for its content.

# VI. APPENDIX

## VI.A. Abbreviations and *Nomenclature*

*Table 5. Abbreviations sorted alphabetically.*

| | |
|---|---|
| AS | Ancillary services |
| BSS | Battery storage system |
| EEG | German renewable energy law |
| ENTSO-E | European Network of Transmission System Operators for Electricity |
| EPR | Energy-to-power ratio |
| EV | Electric vehicle |
| FCR | Frequency containment reserve |
| FNA | (German) Federal Network Agency |
| NACE | European Classification of Economic Activities |
| OCPP | Open Charge Point Protocol |
| PGS | Institute for Power Generation and Storage Systems |
| RWTH Aachen University | Rheinisch-Westfälische Technische Hochschule Aachen |
| SOC | State-of-charge |
| SOE | State-of-energy |
| TSO | Transmission System Operator |
| V2G | vehicle-to-grid |
| V2H | vehicle-to-home |

*Table 6. Nomenclature*

| | |
|---|---|
| $w_{start}$ | trip start probability |
| $P_{EV}$ | Rated power of the battery converter |
| $P_{CS}$ | Rated power of the charging station |
| $P_{charge}$ | Charge power for FCR |
| $P_{discharge}$ | Discharge power for FCR |
| $E_{Bat}$ | Battery energy capacity |
| $E_{market}$ | Marketable energy |
| $E_{mobility}$ | Reserved energy for mobility |
| $E_{charge}$ | Charge energy for FCR |
| $E_{discharge}$ | Discharge energy for FCR |
| $\Delta T_{supply}$ | Duration of FCR service period |

### VI.B. *Additional information on used databases*

*Table 7: EVs from database "Measurements" used in the project GO-ELK [46]. Some EVs were switched between the different trades, which is why the number of EVs is greater than the measured 22 EVs.*

| **Human health** | **Energy provider** | **Carsharing** | | **Privat persons** |
|---|---|---|---|---|
| | | Tech-Park | Rural area | |
| Smart E.D. (17,6 kWh) | Nissan e-NV200 (24 kWh) | Nissan Leaf (24 kWh) | Smart E.D. (17,6 kWh) | Kangoo ZE (22 kWh) |
| Smart E.D. (17,6 kWh) | Nissan Leaf (24 kWh) | Opel Ampera (16,5 kWh) | Nissan Leaf (24 kWh) | Peugeot iOn (16 kWh) |
| Smart E.D. (17,6 kWh) | Nissan Leaf (24 kWh) | BMW i3 (21,6 kWh) | | Peugeot iOn (16 kWh) |
| Mitsubishi i-MiEV (16 kWh) | Smart E.D. (17,6 kWh) | Smart E.D. (17,6 kWh) | | Peugeot iOn (16 kWh) |
| VW e-up! (18,7 kWh) | Kangoo ZE (22 kWh) | Smart E.D. (17,6 kWh) | | BMW i3 Rex (21,6 kWh) |
| Opel Ampera (16,5 kWh) | Nissan Leaf (24 kWh) | Smart E.D. (17,6 kWh) | | |
| Nissan Leaf (24 kWh) | | | | |

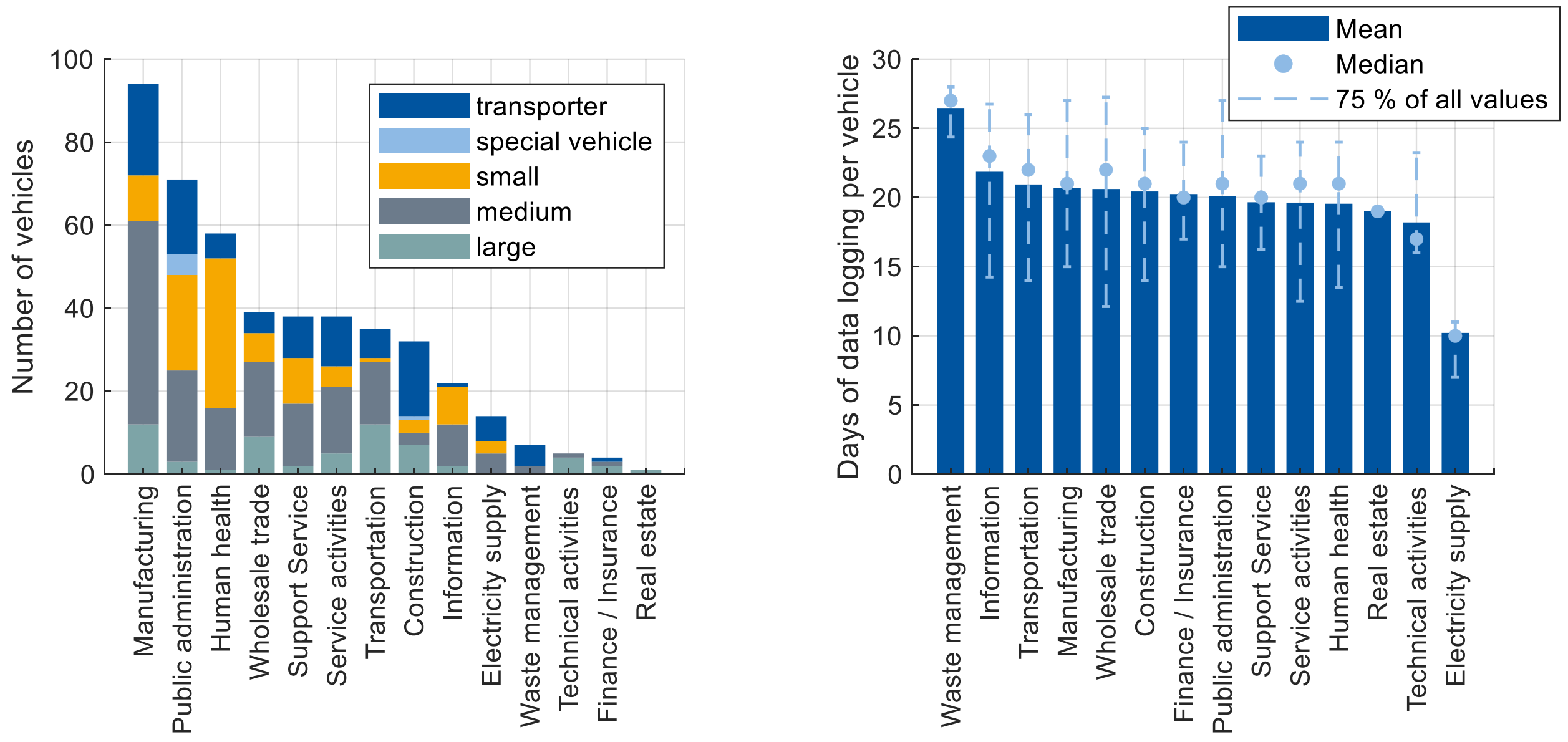

*Figure 21: Vehicles clustered according to economic sector (left) and mean duration of data logging (right) from database "Logbook".*

*Table 8: Electric vehicle data used to assume battery capacity and energy consumption for database "Logbook". Calculation based on data from ADAC [56].*

| Vehicle size | Differentiation in REM2030 according to cubic centimeters (cc) | Assumed Differentiation E-vehicles | Brand & model | Battery Capacity kWh | Consumption kWh/100km | ADAC Real Consumption kWh/100km | Factor between nominal and real consumption | Vehicle weight kg | Torque Nm |
|---|---|---|---|---|---|---|---|---|---|
| **Small** | Displacement < 1,400 cc | Torque < 220 Nm & Weight 1400 kg to 2000 kg | Citroën C-Zero | 14.5 | 12.6 | - | - | 1,440 | 196 |
| | | | Citroën E-Mehari | 30 | 20 | - | - | 1,838 | 166 |
| | | | Peugeot iOn | 14.5 | 12.6 | 16.94 | 1.34 | 1,450 | 180 |
| | | | Renault Zoe (22 kWh) Life | 22 | 13,3 | 21.4 | 1.61 | 1,943 | 220 |
| | | | Smart Fortwo coupé electric drive | 17,6 | 15,1 | 19.2 | 1.27 | 1,150 | 130 |
| | | | Average | 19.10 | 14.52 | 18.89 | 1.39 | 1,545 | 179 |
| | | | **Assumed** | **19.1** | | **18.9** | | | |
| | | | | | | | | | |
| **Medium** | Displacement 1,400 cc to 2,000 cc | Torque 220 Nm to 380 Nm & Weight 1600 kg to 2200 kg | BMW i3 (94 Ah) | 33.2 | 13.1 | 17.4 | 1.33 | 1,620 | 250 |
| | | | Ford Focus Electric | 33.5 | 15.4 | 22.4 | 1.45 | 2,085 | 250 |
| | | | Hyundai Ioniq Elektro | 28 | 11.5 | 14.7 | 1.28 | 1,880 | 295 |
| | | | KIA Soul EV | 30 | 14.7 | 19.4 | 1.32 | 1,960 | 285 |
| | | | Mercedes-Benz B 250 e | 28 | 16.6 | 20.2 | 1.22 | 2,170 | 340 |
| | | | Nissan Leaf | 24 | 15 | 20.39 | 1.36 | 1,965 | 280 |
| | | | Opel Ampera-E | 60 | 14.5 | 19.7 | 1.36 | 2,056 | 360 |
| | | | VW eGolf | 24.2 | 12.7 | 18.2 | 1.43 | 1,960 | 270 |
| | | | Volvo C30 Electric | 22.7 | 15 | 28.3 | 1.89 | 1,995 | 220 |
| | | | Average | 31.36 | 14.35 | 20.08 | 1.40 | 1,966 | 283 |
| | | | **Assumed** | **31.4** | | **20.1** | | | |
| | | | | | | | | | |
| **Large** | Displacement 1,400 cc to 2,000 cc | Torque 220 Nm to 380 Nm & Weight 1600 kg to 2200 kg | Audi e-tron 55 quattro | 95 | 23 | - | - | 2,565 | 664 |
| | | | BMW Concept ix3 (2020) | 70 | 17.5 (calc) | - | - | | 561 |
| | | | Hyundai Kona Elektro | 39.2 | 14.3 | - | - | 1,760 | 395 |
| | | | Jaguar I-Pace | 90 | 21.2 | 27.6 | 1.32 | 2,208 | 696 |
| | | | Tesla Model S P90D | 90 | 17.8 | 24 | 1.35 | 2,670 | 967 |
| | | | Tesla Model X | 100 | 20.8 | 24 | 1.15 | 2,534 | 660 |
| | | | Average | 80.7 | 19.31 | 24 | 1.27 | - | 680 |
| | | | **Assumed** | **80.7** | | **27** [2] | | | |
| | | | | | | | | | |
| **Trans-porter** | Displacement > 1,400 cc Weight < 3,500 kg | Weight 1,644 kg to 2,600 kg & mostly 2-3 seats with a lot of storage size | Citroen Berlingo Electric L2 | 22.5 | 17.7 (NEFZ) | - | - | 1,644 | 200 |
| | | | Iveco Daily Electric | 28.2 | - | - | - | 2,500 | 300 |
| | | | Nissan e-NV200 | 24 | 16.5 | 22.8 | 1.38 | 1.640 | 254 |
| | | | Peugeot Partner Electric | 22.5 | 17.7 | - | - | 1,664 | 152 |
| | | | Renault Kangoo Z.E. | 22 | 15.5 | 23.5 | 1.52 | 1,520 | 226 |
| | | | Streetscooter Work L Box | 40 | 19.2 (NEFZ) | - | - | 1,640 | 200 |
| | | | VW eCrafter | 35.8 | 21.54 | - | - | 2,522 | 290 |
| | | | Average | 27.86 | 18.02 | 23.15 | 1.45 | 2,158 | 232 |
| | | | **Assumed** | **27.9** | | **25.2** [2] | | | |

[2] As only few vehicles were tested by the ADAC, the real electricity consumption for large vehicles and transporter is calculated using the factor 1.4 (see small and medium average factor) and multiply it with the average nominal consumption.

### VI.C. Literature for further research

*Table 9: Summary of **projects** working on the provision of ancillary services using EV fleets*

| Source | Date | Name | Partner | Focus & Results |
|---|---|---|---|---|
| [28] | 2002 | "Vehicle-to-Grid Demonstration Project: Grid Regulation Ancillary Service with a Battery Electric Vehicle" (V2GDP) | AC Propulsion, California Air Resources Board, California Environmental Protection Agency, | - Evaluation of the feasibility of the provision of grid regulation using EV<br>- EV are able to provide grid regulation and the ISO system requirements regarding data transmission times could be fulfilled<br>- Energy throughput when providing regulation power is equivalent to that resulting from daily driving |
| [30] | 2012 - 2015 | "Intelligente Netzanbindung von Elektrofahrzeugen zur Erbringung von Systemdienstleistungen – INEES" (Intelligent grid integration of EV to provide system services) | Fraunhofer IWES, LichtBlick SE, SMA Solar Technology AG, Volkswagen AG | - Field tests of the provision of secondary control reserve using a fleet of 20 V2G-capable EV<br>- Provision is technically possible, but under current costs and revenue not profitable |
| [32] | 2013 - 2018 | Los Angeles Air Force Base Vehicle to Grid Demonstration (LAAFB) | Lawrence Berkeley National Laboratory (LBNL), Kisensum LLC | - Demonstration of a fleet of 29 bidirectional EV providing frequency regulation to generate revenue<br>- Charging stations and EV should have a capacity/power ratio of at least two to participate in a fleet offering frequency regulation |
| [34] | 2016 - 2019 | The Parker Project (Parker) | DTU, Nuvve, Nissan, Insero, Enel X, Groupe PSA, Mitsubishi Corporation, Mitsubishi Motors Corporation, Frederiksberg Forsyning | - Demonstration project to analyze the integration of V2G-capable EV into the electricity grid<br>- Results show that EV are able to provide ancillary services<br>- Recommendations are the planning of electrification of transportation, continuous research, "test zones and pilots on new market designs" and an international collaboration |
| [41] | 2018 - 2019 | Industrial Pilot Project | The Mobility House, ENERVIE, Amprion, Nissan | - Demonstration of the provision of FCR using one EV that got prequalified from the German TSO |
| [42] | 2019 - 2021 | "Bidirectional Charging Management – Field Trial and Measurement Concept for Assessment of Novel Charging Strategies" | BMW, FfE e.V., FfE GmbH, Kostal Industrie Elektrik GmbH, TenneT TSO GmbH, Bayernwerk Netz GmbH, Karlsruhe Institute of Technology (KIT), University Passau | - Analysis of the interaction between EV, charging infrastructure and the power grid<br>- Identification and demonstration (using 50 EV) of use-cases of V2G in different markets. |
| [44,45] | 2019 - 2021 | Industrial Pilot Project | Tennet, Next Kraftwerke, Jedlix | - Field test of EV providing frequency regulation in a virtual power plant<br>- Customers of Jedlix charging their EV receive financial benefit when providing secondary control reserve |

*Table 10: Summary of literature about* ***demonstrations, experiments and field tests*** *of the provision of frequency regulation using EV fleets*

| Source | Date | Project | Focus | Results |
|---|---|---|---|---|
| Brooks, Gage [29] | 2001 | V2GDP | Analysis of ancillary services EV, hybrid vehicles and fuel-cell vehicles may provide by showing test results | - Field tests show that the EV is capable of providing power and thus benefit to the grid<br>- EV might be able to achieve lower net ownership costs in comparison to conventional vehicles by providing grid services |
| Marinelli et al. [35] | 2016 | Parker | Centralized approach to provide FCR with EV using unidirectional charging and experimental validation of the approach | - Provision of FCR with EV by only using unidirectional charging is viable with fast response time |
| Thingvad et al. [36] | 2016 | Parker | Economic comparison of EV fleet providing Frequency Normal-operation Reserve (FNR) through unidirectional vs. bidirectional (V2G) charging in Eastern Denmark | - Bidirectional FNR is more lucrative (factor of 6.6-13.3) and viable than unidirectional as it can be applied longer and independently of the driven distance<br>- Experiments show that EV are able to perform unidirectional FNR and bidirectional FNR with delay times of 1 respectively 5 seconds |
| DeForest et al. [33] | 2017 | LAAFB | EV fleet participating in California Independent System Operator (CASIO) frequency regulation market | - Development of a Day-Ahead optimization model applied to the Los Angeles Air Force Base EV fleet minimizing operation cost and maximizing revenue from ancillary service |
| Degner et al. [31] | 2017 | INEES | Analysis of the effects of EV secondary control reserve provision on the distribution grid using simulations and field tests | - Power quality of the distribution grid is not negatively influenced by the EV provision<br>- The EV impact on the distribution grid can be anticipated and managed well |
| Hashemi et al. [37] | 2018 | Parker | Presentation of results from three different EV (Nissan Leaf, Peugeot iOn and Mitsubishi Outlander) providing FCR-N (frequency-controlled normal operation reserve) in Nord Pool energy market | - All three EV were able to response within five seconds and with an accuracy of around 98%<br>- The depth-of-discharges (DoDs) were always smaller than 40% |
| Bañol Arias et al. [38] | 2018 | Parker | Analysis of an EV fleet participating in Danish FNR market and determination of issues appearing in the field | - EV fleet is able to support the grid, e.g. in FNR market, but availability of the EV is crucial<br>- Practical issues are "communication delays, measurement errors and physical equipment constraints" [38] |
| Thingvad et al. [39] | 2019 | Parker | Economic analysis of EV performing FNR in the Nordic countries considering requirements and losses | - The value of bidirectional FNR is much higher than providing unidirectional FNR<br>- Suggestion: Aggregator should pay for the EV's driving energy consumption to remunerate owners<br>- Losses due to charger's efficiency results in reduced revenue by 22% |
| Bañol Arias et al. [40] | 2020 | Parker | Economic assessment of EV participating in FCR-N in the Nord Pool market from the owner's perspective applying three operation strategies: complete pause, over-fulfillment, preferred operating point | - EV owners can achieve profits from the provision of FCR-N, mostly when choosing a smart operation strategy<br>- Preferred operating point strategy resulted in highest profits (up to 1100€ per EV per year when bidding 10 kW) as it minimized the battery degradation and the unavailability times |

*Table 11: Summary of selected literature of German rules and regulations on the frequency containment reserve (FCR) market*

| Source | Date | Content related to the provision of FCR using battery energy storage systems |
|---|---|---|
| VDN [77] | 2007 | TransmissionCode 2007: Network and system rules of the German transmission system operators<br>In Appendix D: Documents for prequalification for the Provision of primary control power to TSO - Degrees of freedom, rules and requirements that must be met by a provider of FCR |
| FNA [61] | 2011 | German federal network agency (FNA) changes FCR bidding time from monthly to weekly |
| German TSO [76] | 2014 | Key points and degrees of freedom for the provision of FCR using BSS as an example |
| German TSO [67] | 2015 | Storage capacity requirements for the provision of FCR using batteries (e.g. 30-minute-criterion) |
| FNA [63] | 2018 | German federal network agency (FNA) changes FCR bidding time from weekly to daily starting in July 2019 and from daily to 6 daily sections of 4 hours starting in July 2020. |
| FNA [11] | 2019 | Decision of the German federal network agency (FNA) to stop the TSO from requiring the 30-minute-criterion when providing FCR with BSS. From now on 15-minute-criterion for all providers including BSS. |
| German TSOs [69] | 2019 | Minimal requirements on the IT when providing control reserve. When pooling small systems (< 25 kW per system, maximum pool size 2 MW) the connection between the systems can from now on be made via the internet. |